\documentclass[prd,twocolumn,showpacs,amsmath,amssymb,preprintnumbers,nofootinbib,longbibliography,showkeys]{revtex4-1}
\usepackage{graphicx}
\usepackage{amsmath}
\usepackage{amssymb}
\usepackage{xspace}
\usepackage{gensymb}
\usepackage{hhline}
\usepackage{multirow}
\usepackage{booktabs}
\usepackage{hyperref}
\usepackage{cleveref}
\usepackage{aas_macros}
\usepackage{soul}
\usepackage{xcolor}
\usepackage{array}
\usepackage{adjustbox}
\usepackage{cancel}

\newlength{\wdo}
\newcommand{\stroke}[1]{{$#1$}%
\settowidth{\wdo}{${#1}$} {\kern-\wdo}%
\partialvartstrokedint}

\makeatletter
\newcommand{\fancysep}{%
  \@afterindentfalse
  {\begin{center}
    \resizebox{0.8\linewidth}{0.4ex}{{%
        \fontsize{20}{24}\usefont{U}{webo}{xl}{n}{4}}}%
  \end{center}}\@afterheading}
\makeatother

{\vskip 1.0em
\framebox[\columnwidth][r]{%
\begin{minipage}[c]{\columnwidth}%
\vspace{-1.0em}%
#1%
\end{minipage}}}
{\vskip 1.0em}

\def\XXint#1#2#3{{\setbox0=\hbox{$#1{#2#3}{\int}$}
     \vcenter{\hbox{$#2#3$}}\kern-.5\wd0}}

\newcommand{\beq}{\begin{equation}}
\newcommand{\eeq}{\end{equation}}
\newcommand{\beqa}{\begin{eqnarray}}
\newcommand{\eeqa}{\end{eqnarray}}

\newcommand{\Heiii}{\ensuremath{{^3}\textnormal{He}}\xspace}

\newcommand{\neff}{\ensuremath{N_\textnormal{eff}}\xspace}

\newcommand{\phie}{\ensuremath{\phi_e}\xspace}

\newcommand{\burst}{{\sc burst}\xspace}

\newcommand{\tcm}{\ensuremath{T_\textnormal{cm}}\xspace}

\newcommand{\np}{\ensuremath{n/p}\xspace}
\newcommand{\nprates}{\ensuremath{n\leftrightarrow p}\xspace rates\xspace}

\newcommand{\nue}{{\ensuremath{\nu_{e}}}\xspace}

\newcommand{\bnue}{\ensuremath{\overline{\nu}_e}\xspace}

\newcommand{\ben}{\begin{enumerate}}
\newcommand{\een}{\end{enumerate}}

\newcommand{\heiv}{\ensuremath{\,^4\textnormal{He}}\xspace}

\newcommand{\livii}{\ensuremath{{}^7\textnormal{Li}}\xspace}
\newcommand{\livi}{\ensuremath{{}^6\textnormal{Li}}\xspace}
\newcommand{\bevii}{\ensuremath{{}^7\textnormal{Be}}\xspace}
\newcommand{\yp}{\ensuremath{Y_P}\xspace}

\newcommand{\mesa}{{\sc mesa}\xspace}
\newcommand{\deltamnp}{\ensuremath{\delta m_{np}}\xspace}

\newcommand{\hi}{\ensuremath{{}^1\textnormal{H}}\xspace}

\newcommand{\hethree}{\ensuremath{\,^3\textnormal{He}}\xspace}

\newcommand{\osixteen}{\ensuremath{\,^{16}\textnormal{O}}\xspace}

\newcommand{\carbonxii}{\ensuremath{{}^{12}\textnormal{C}}\xspace}
\newcommand{\oxvi}{\ensuremath{{}^{16}\textnormal{O}}\xspace}

\newcommand{\teff}{\ensuremath{T_{\rm eff}}\xspace}

\newcommand{\be}{\begin{equation}}
\newcommand{\ee}{\end{equation}}
\newcommand{\simless}{\lower.5ex\hbox{$\; \buildrel < \over \sim\;$}}
\newcommand{\simgreat}{\lower.5ex\hbox{$\; \buildrel > \over \sim\;$}} 

\newcommand{\sigmanp}{\langle \sigma v \rangle_{\rm np}}

\newcommand{\mpro}{\ensuremath{m_{\rm p}}\xspace} 
 
\newcommand{\epcool}{\epsilon_{\rm cool}}
\newcommand{\epnuke}{\epsilon_{\rm nuke}}

\begin{document}

\title{Universes without the Weak Force:
Astrophysical Processes with Stable Neutrons}

\author{E. Grohs$^{1}$}
\author{Alex R. Howe$^2$}
\author{Fred C. Adams$^{1,2}$}

\affiliation{$^{1}$Department of Physics, University of Michigan, Ann Arbor,
Michigan 48109, USA}
\affiliation{$^{2}$Department of Astronomy, University of Michigan, Ann Arbor,
Michigan 48109, USA}

\date{\today}

\begin{abstract}

We investigate a class of universes in which the weak interaction is not in
operation.  We consider how astrophysical processes are altered in the absence
of weak forces, including Big Bang Nucleosynthesis (BBN), galaxy formation,
molecular cloud assembly, star formation, and stellar evolution. Without weak
interactions, neutrons no longer decay, and the universe emerges from its early
epochs with a mixture of protons, neutrons, deuterium, and helium. The
baryon-to-photon ratio must be smaller than the canonical value in our universe
to allow free nucleons to survive the BBN epoch without being incorporated into
heavier nuclei. At later times, the free neutrons readily combine with protons
to make deuterium in sufficiently dense parts of the interstellar medium, and
provide a power source before they are incorporated into stars.  Almost all of
the neutrons are incorporated into deuterium nuclei before stars are formed. As
a result, stellar evolution proceeds primarily through strong interactions,
with deuterium first burning into helium, and then helium fusing into carbon.
Low-mass deuterium-burning stars can be long-lived, and higher mass stars can
synthesize the heavier elements necessary for life.  Although somewhat
different from our own, such universes remain potentially habitable.

\end{abstract}



\maketitle

\section{Introduction}\label{sec:intro} 

The fundamental constants that describe the laws of physics appear to have
arbitrary values that cannot be explained by current theory. One possible ---
and partial --- explanation is that other universes exist in which the
fundamental constants have different values, so that they are drawn from an
as-yet-unknown probability distribution. Many authors have argued that
significant changes in these constants could render the universe uninhabitable
to life as we know it, and as a result our universe appears to be
``fine-tuned'' for life \cite{bartip,carr,carter,hogan,reessix}.  On the other
hand, recent work suggests that when multiple constants are allowed to vary,
large regions of the parameter space that result in habitable universes can be
found \cite{adams2008,adams2016,coppess}. This paper continues the exploration
of alternate possibilities for the fundamental constants with a focus on the
weak force. 

The strength of the weak interaction is a fundamental part of the standard
model of particle physics and represents one parameter that could vary from
region to region. The weak interaction governs the rate of radioactive decay,
the rate of conversion of hydrogen into helium via the $p(p,\nu_e e^+){\rm D}$
stage of the $pp$-chain in low-mass stars, and the cross section for neutrino
interactions.  The latter two effects are crucial to determining whether or not
a universe can produce life. If the weak force is too weak, or absent
altogether, long-lived stars fueled by weak reactions could not exist. In the
absence of weak interactions, helium can still be synthesized through strong
interactions during Big Bang Nucleosynthesis (BBN), as free neutrons are
present, but helium production is suppressed in stellar interiors (which do not
have neutrons).  Once synthesized, helium can later fuse into heavier elements
via the triple alpha process in sufficiently massive stars.  However, without
neutrino interactions, core-collapse supernova will fail to explode and will
simply collapse to a degenerate remnant, thereby hampering the dispersal of
heavy elements.

These effects would appear to compromise the habitability of universes where
the weak force is significantly weaker than in our own.  However, this standard
argument has been challenged with the concept of a ``weakless universe''
\cite{HKP:2006_weakless}, namely, a universe without any weak interaction at
all. Neutrons would be stable in such a universe, and would therefore most
likely have an equal abundance to protons. A universe with the same baryon
density as ours would convert virtually all baryons to helium during BBN,
leaving behind no free protons (no hydrogen to make water). However, if the
baryon density is lower (more properly the baryon-to-photon ratio $\eta$),
significant amounts of protons and deuterium can survive the BBN epoch
\cite{HKP:2006_weakless} and would be available later for deuterium burning in
stars (which takes place through the strong interaction). Note that stars in
such a weakless universe can also produce heavier elements by strong reactions.
Although core-collapse supernova might not function, at least not in the same
manner as in our universe, these heavy elements could still be dispersed by
Type Ia supernova and classical novae, allowing the possibility of planet
formation and life.

The opposite case, where the weak interaction is significantly stronger than in
our universe, would resemble our universe more closely. A stronger weak
interaction increases the rate of radioactive decay, most notably the neutron
lifetime, but it does not affect the stability of stable nuclei, which is
governed primarily by the strong interaction. If the neutron lifetime is less
than 30 seconds, BBN is suppressed because most neutrons decay before BBN
begins. However, this complication is not an impediment because stars in this
all-hydrogen universe will be more efficient at converting hydrogen to helium.
In addition, core-collapse supernovae will be more efficient at dispersing
heavy elements due to the larger neutrino interaction cross sections. The
primary concern would be the shortened lifetimes of these more efficient stars.
(Note that if the weak interaction is strong enough to approach the strength of
the electromagnetic interaction, non-linear effects will likely render these
concerns moot.)

In this paper, we update and expand upon this idea of a weakless universe in
Refs.\ \cite{2006hep.ph....9050C,2011PhRvD..83k5020G,2012arXiv1207.4861G}, and
in particular Ref.\ \cite{HKP:2006_weakless}.  An important parameter in this
problem is the baryon-to-photon ratio $\eta$, which impacts the composition of
the universe after BBN. For the value in our universe, BBN with an equal amount
of neutrons and protons would result in a composition where the $^4$He
abundance is greater than $90\%$.  The high abundance of \heiv yields
short-lived stars and is problematic for the development of life. For a
judicious choice of $\eta$, however, BBN can result in a richer composition
with large fractions of deuterium, free protons, and free neutrons.  We
determine the range of $\eta$ for which weakless universes could support life,
using models of galaxy formation, star formation, and subsequent stellar
evolution.

One crucial issue not addressed in the original proposal
\cite{HKP:2006_weakless} is the abundance of free neutrons left over from BBN,
which is comparable to the abundance of free protons. These neutrons can
capture onto protons at zero temperature, forming deuterium via the
$n(p,\gamma){\rm D}$ reaction.  As a result, nuclear fusion can occur in the
interstellar medium and could potentially halt the collapse of a gas cloud.
This paper considers the effects of neutron capture reactions on four scales of
formation: [A] from the intergalactic medium down to the size scale of
galaxies; [B] from the neutral interstellar medium (ISM) to the formation of
giant molecular clouds; [C] from the cloud to molecular cloud cores -- the
sites of individual star formation events; and finally [D] from the cloud cores
to the production of protostars themselves.  Neutron fusion becomes significant
during the latter stages of this hierarchy, and most of the free neutrons are
processed into deuterium before the onset of stellar nuclear burning.  Although
gas cooling is delayed, these processes do not disrupt star formation entirely.

We next consider stellar evolution and nuclear burning in a weakless universe
and the chemical evolution of the universe through successive generations of
stars. Long-lived stars can exist in a weakless universe if the deuterium
abundance is high enough for deuterium fusion to continue on Gyr timescales.
With the main sources of heavy element dispersal being red giant winds and Type
Ia supernovae, the composition of the interstellar medium will be different
from our universe, relatively enriched in carbon and iron peak elements and
depleted in oxygen.  A low oxygen abundance could cause a dearth of water which
would impose a problem to life -- assuming water is an essential ingredient to
life like it is on our planet.  However, later generations of stars will
undergo different reactions that will likely mitigate this problem.

Iron peak elements in the ISM will capture an excess of free neutrons, allowing
stars to form with a small excess of protons, which can then form oxygen via
the reaction $\carbonxii(2p,\gamma){}^{14}{\rm O}$. Similar reactions may lead to a
deficit of nitrogen and an excess of neon compared with our universe, and
likewise, the lack of core-collapse supernovae will likely lead to a deficit of
non-alpha-process elements. However, the effect on the chemical evolution of
the weakless universe is neither negligible nor dominant, and the necessary
elements for both planet formation and organic chemistry will still be present,
implying that such a universe could still be hospitable to life.

The organization of this paper is as follows.  Section \ref{sec:bbn} details
the outcome of BBN in universes with a range of $\eta$ and neutron-to-proton
ratios.  We examine the impact of weakless physics on: galaxy formation in
Sec.\ \ref{sec:galaxy}; the ISM in Sec.\ \ref{sec:ism}; and stellar evolution
in Sec.\ \ref{sec:stars}.  Section \ref{sec:ChemEvo} contains our study of the
habitability of the weakless universe.  We summarize and discuss our results in
Sec.\ \ref{sec:conclude}.  Throughout this paper, we use cgs units with the
exception of BBN in Sec.\ \ref{sec:bbn}.  We use natural units in Sec.\
\ref{sec:bbn} to be consistent with the BBN literature.

\section{Big Bang Nucleosynthesis}\label{sec:bbn} 

\subsection{Standard versus weakless BBN}

In this section, we give a brief overview of the role the weak interaction
plays in Standard BBN (SBBN) and then discuss the modifications to SBBN in a
weakless universe.  We will use the ratio of neutrons to protons (denoted the
\np ratio) extensively throughout this work
\beq\label{eq:np_def}
  \np\equiv\frac{n_n}{n_p}.
\eeq
Note that the number densities $n_i$ in Eq.\ \eqref{eq:np_def} are the {\it
total} particle number densities (free particles and nucleons bound in nuclei)
and not solely the free particle number densities.  Reference
\cite{1950PThPh...5..224H} showed that \np would evolve due to six weak
reactions during primordial nucleosynthesis. We schematically write the six
reactions as
\begin{align}
  \nu_e + n&\leftrightarrow p + e^-\label{eq:np1},\\
  e^+ + n&\leftrightarrow p + \overline{\nu}_e\label{eq:np2},\\
  n&\leftrightarrow p + e^- + \overline{\nu}_e.\label{eq:np3}
\end{align}
and colloquially refer to them as the \nprates.  In the approximation that the
nucleon rest masses are much heavier than both the neutrino and electron
masses, Ref.\ \cite{WFO_approx} gives the prescription for calculating the
rates for each of the six reactions listed in Eqs.\ \ref{eq:np1} --
\ref{eq:np3}.  At high temperature $T$, the weak interaction rates are fast and
maintain the \np ratio in weak equilibrium as a function of $T$
\beq\label{eq:np_evolution}
  \np\,(T) = \exp\left(\frac{-\delta m_{np}+\mu_{e^-}-\mu_{\nu_e}}{T}\right),
\eeq
where $\delta m_{np}=1.293\,\,{\rm MeV}$ is the mass difference between a
neutron and proton, $\mu_{e^-}$ is the chemical potential of the electrons, and
$\mu_{\nu_e}$ is the chemical potential of the electron neutrinos.  In SBBN,
the chemical potentials of the electron and electron neutrino are small and so
the mass term dominates.  Therefore, $\np\simeq1$ at high temperature.  Figure
6 of Ref.\ \cite{WFO_approx} shows how \np initially follows an equilibrium
track for temperatures above $1\,\,{\rm MeV}$, and diverges from equilibrium at
lower temperature.  Free neutron decay [the forward process in Eq.\
\eqref{eq:np3}] would eventually transmute all neutrons into protons if there
were no other nuclear processes.  \np would go to zero and the universe would
emerge from BBN with a pure \hi composition.  Such a scenario does not
transpire as a chain of reactions assembles free protons and neutrons into
\heiv and some other low atomic mass nuclides.  Nuclear freeze-out of \heiv
occurs when the temperature has reached $T\sim100\,\,{\rm keV}$ and
$\np\sim1/7$.

In a weakless universe, there are no weak interactions and Eqs.\ \eqref{eq:np1}
-- \eqref{eq:np3} do not apply.  This implies that neutrons are stable to
decay.  The neutron to proton ratio is set at a high temperature and is a
function of the ratio of up quarks to down quarks, $u/d$
\beq\label{eq:np_ud}
  \np = \frac{2-u/d}{2u/d-1}.
\eeq
$u/d$ ranges from $1/2$ for a pure neutron universe to $2$ for a pure proton
universe.  Assuming baryon number conservation, \np is fixed and does not
evolve.  Weakless BBN (wBBN) proceeds with a fixed \np ratio.  If there are
multiple families of quarks with nonzero densities, then there would be other
hadrons present.  We do not consider such scenarios in this work.

\subsection{Description of calculations}

We employ the code \burst \cite{GFKP-5pts:2014mn} to do both SBBN and wBBN
calculations.  \burst is based off of the work in Refs.\ \cite{WFH:1967} and
\cite{SMK:1993bb}.  In a SBBN calculation, there are four sets of quantities to
evolve as a function of time.  There are three thermodynamic variables: the
plasma temperature $T$; the scale factor $a$; and the electron-degeneracy
parameter \phie where
\beq
  \phie\equiv\frac{\mu_{e^-}}{T}.
\eeq
All thermodynamic quantities of the plasma, such as energy density and
pressure, are functions of $T$, $a$, and \phie.  The last set of evolution
quantities are the abundances $Y_i$.  Abundances are related to the mass
fractions $X_i$ via
\beq
  Y_i = \frac{X_i}{A_i},
\eeq
where $A_i$ is the atomic mass number for species $i$.  To compare SBBN and
wBBN calculations, we will employ a network of nine nuclides which include all
bound nuclides up to mass number $A=7$.  The nuclear reaction network includes
34 strong, electromagnetic, and weak interactions from Refs.\ \cite{SMK:1993bb}
and \cite{CF:88}.  For more precise calculations of SBBN, one can evolve the
neutrino spectra as they go out of equilibrium which can lead to changes at the
$1\%$ level \cite{Trans_BBN}.  The present work does not require this level of
precision, so we always assume neutrinos are in a Fermi-Dirac distribution with
zero chemical potential and a temperature given by the comoving temperature
parameter \tcm as a function of scale factor $a$
\beq\label{eq:tcm}
  \tcm(a) \equiv T_{\rm in}\frac{a_{\rm in}}{a}.
\eeq
The subscript ``in'' on the temperature and scale factor symbols denotes an
initial epoch when the neutrinos are in thermal equilibrium with the
electromagnetic components of the plasma.  We must use \tcm because it is
manifestly different from $T$ and has ramifications on the \np ratio
\cite{xmelec}.

A wBBN calculation proceeds in the same way as a SBBN one.  We evolve $T$, $a$,
\phie, and the nine nuclides as a function of time.  We maintain all strong and
electromagnetic cross sections unaltered from SBBN.  There are two key
differences between wBBN and SBBN.  In a weakless universe, there are no weak
nuclear interactions.  We remove the \nprates, i.e., Eqs.\ \eqref{eq:np1} --
\eqref{eq:np3}, and any other $\beta$-decay rates, most notably that of the
three-nucleon hydrogen isotope tritium (denoted as ${\rm T}$), ${\rm
T}\rightarrow\hethree+e^-+\bnue$.  Tritium has a mean lifetime
$\tau_3\sim20\,{\rm years}$ and will not decay into \hethree until well after
SBBN concludes.  If doing precision SBBN calculations, we would add the
abundance of ${\rm T}$ to that of \hethree to compare to observation
\cite{BRB:He3}.  In a weakless universe however, the difference in atomic
number between {\rm T} and \hethree could be important in the initial stages of
stellar evolution.  As a result, we delineate the freeze-out ${\rm T}$ and
\hethree abundances in order to use them in our stellar calculations.

The other difference between SBBN and wBBN is the existence of neutrinos.  Our
model of SBBN in Ref.\ \cite{GFKP-5pts:2014mn} assumes that neutrinos only
interact via the weak interaction and gravity.  We do not know the mechanism
which populates the neutrino states at early times in our universe.  In this
work, we assume that if there is no weak interaction, then there would be no
relic neutrino seas.  We note that if there were neutrinos in a weakless
universe, then the neutrino energy states would not necessarily be thermally
populated.  Alternatively, one could consider the Cosmic Microwave Background
(CMB) observable \neff.  \neff is the ``effective number of neutrino species''
[see Eq.\ (3.11) in Ref.\ \cite{cmbs4_science_book}].  In our universe,
$\neff\sim 3$ for the three flavors of neutrino species \cite{PlanckXIII:2015}.
In a weakless universe where neutrinos were present, \neff would most likely be
smaller than 3.  The exact number would depend on the mechanism which created
the neutrinos.  The neutrinos would behave as a dark radiation component
\cite{dark_rad:2000} and wBBN would proceed with an expansion rate different
than SBBN.  For the wBBN calculations in this work, $\neff=0$.

\subsection{Mass fraction evolution}

In this section we discuss the evolution of the mass fractions as a function of
the \tcm variable from Eq.\ \eqref{eq:tcm}.  For our model of SBBN, the baryon
number is the only cosmological input.  We do not consider other cosmological
inputs such as \neff or neutrino chemical potentials.  There are many
equivalent ways to represent baryon number in BBN.  We adopt the nomenclature
of Ref.\ \cite{SMK:1993bb} and use the baryon-to-photon ratio $\eta$
\beq\label{eq:eta}
  \eta\equiv\frac{n_b}{n_\gamma},
\eeq
where $n_b$ and $n_\gamma$ are the proper number densities of baryons and
photons, respectively.  Based on the strict definition in Eq.\ \eqref{eq:eta},
$\eta$ decreases as electrons and positrons annihilate to become photons for
$T\lesssim100\,{\rm keV}$.  Therefore, we will refer to numerical values of
$\eta$ after the epoch of electron-positron annihilation.  In wBBN, $\eta$ and
the \np ratio are both inputs.  When comparing SBBN to wBBN, we will use the
same value of $\eta$.  There is no way to compare SBBN to wBBN using the same
\np ratio as \np evolves in SBBN according to the $n\leftrightarrow p$ rates
schematically shown in Eqs.\ \eqref{eq:np1} -- \eqref{eq:np3}.  We will pick a
value of the \np ratio to elucidate specific points on how wBBN differs from
SBBN.

\begin{figure}
  \begin{center}
    \includegraphics[width=\columnwidth]{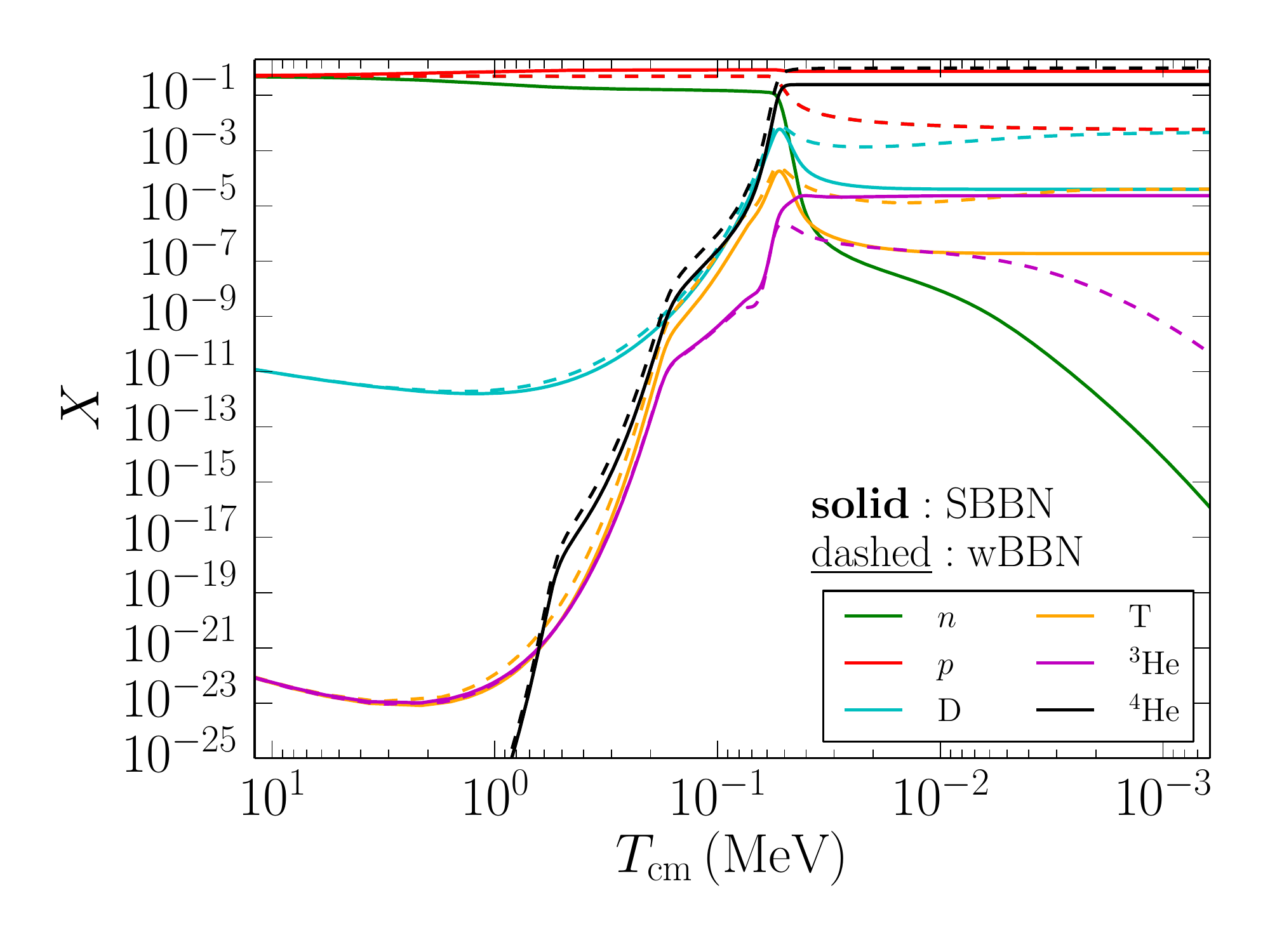}
    \includegraphics[width=\columnwidth]{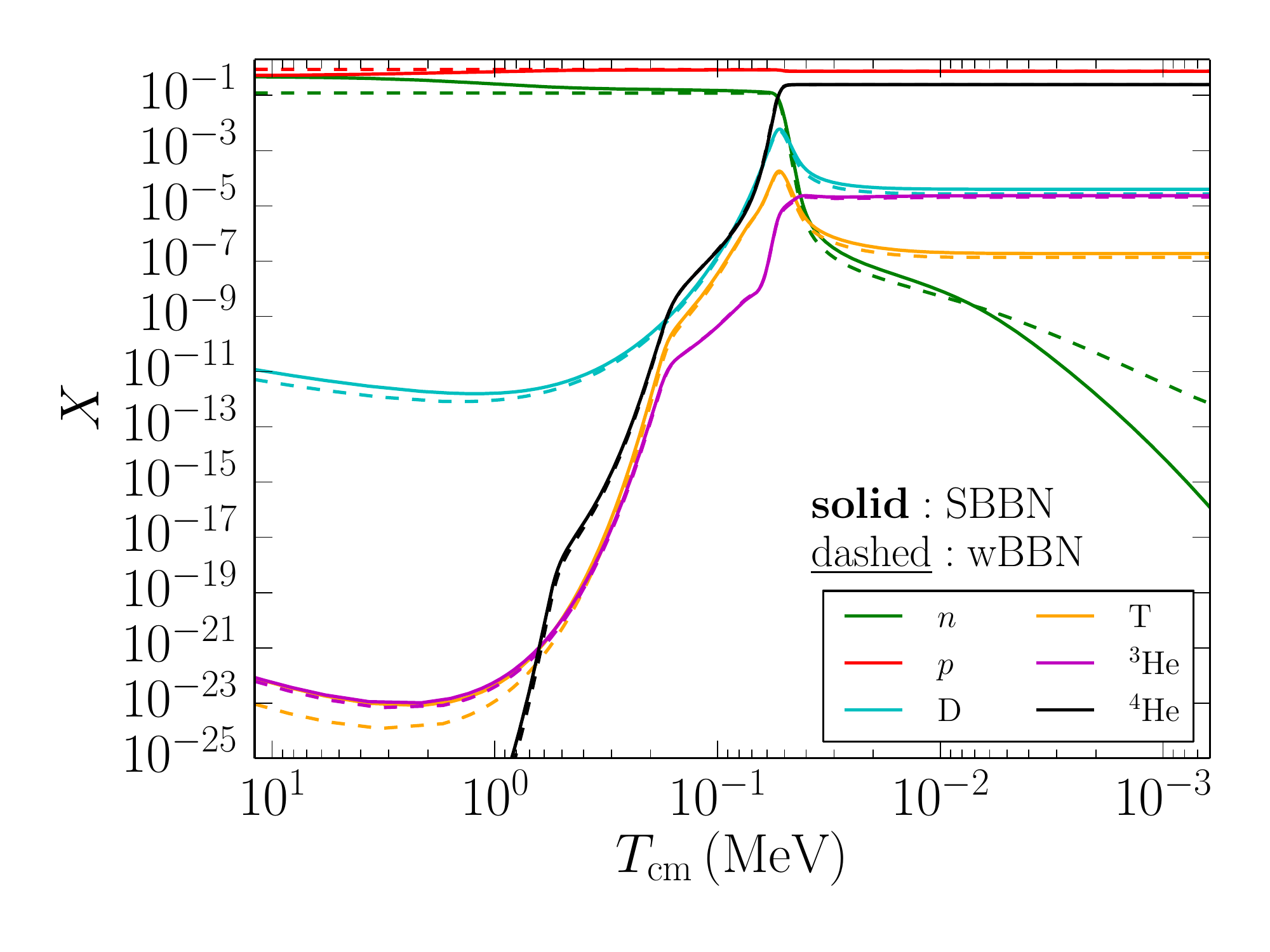}
  \end{center}
  \caption{\label{fig:comp_s1} Mass fractions, $X$, as a function of \tcm.
  Each color represents a different isotope.
  (Top) Solid lines are a SBBN calculation, and dashed lines are a wBBN
  calculation where $\np=1$.  The dashed green line for $n$ is coincident with
  the dashed red line for $p$.  $\eta=6.075\times10^{-10}$ for both calculations.
  (Bottom) Same as the top panel except $\np=0.14$ for the wBBN calculation.
  }
\end{figure}

\begin{figure}
  \begin{center}
    \includegraphics[width=\columnwidth]{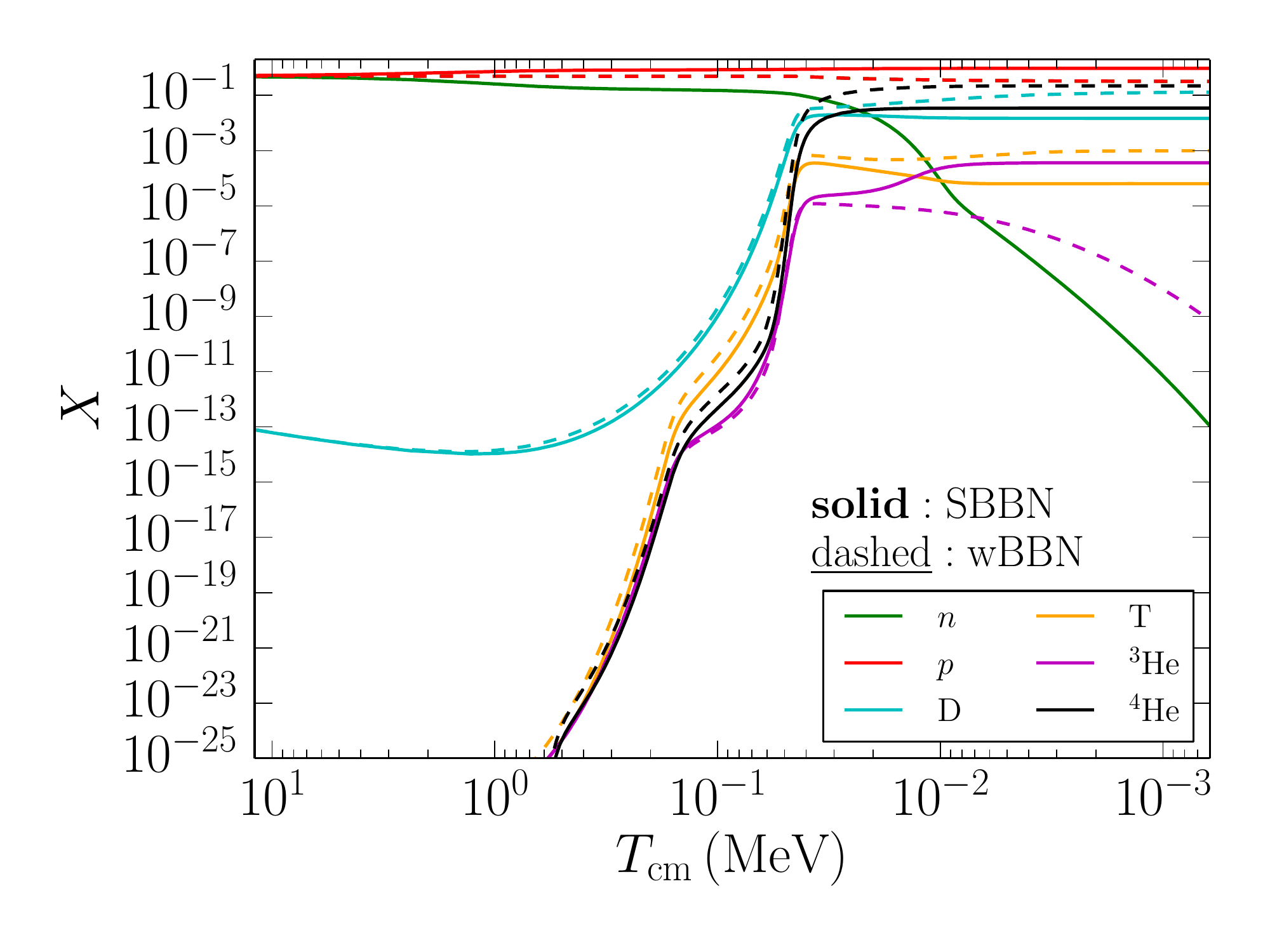}
    \includegraphics[width=\columnwidth]{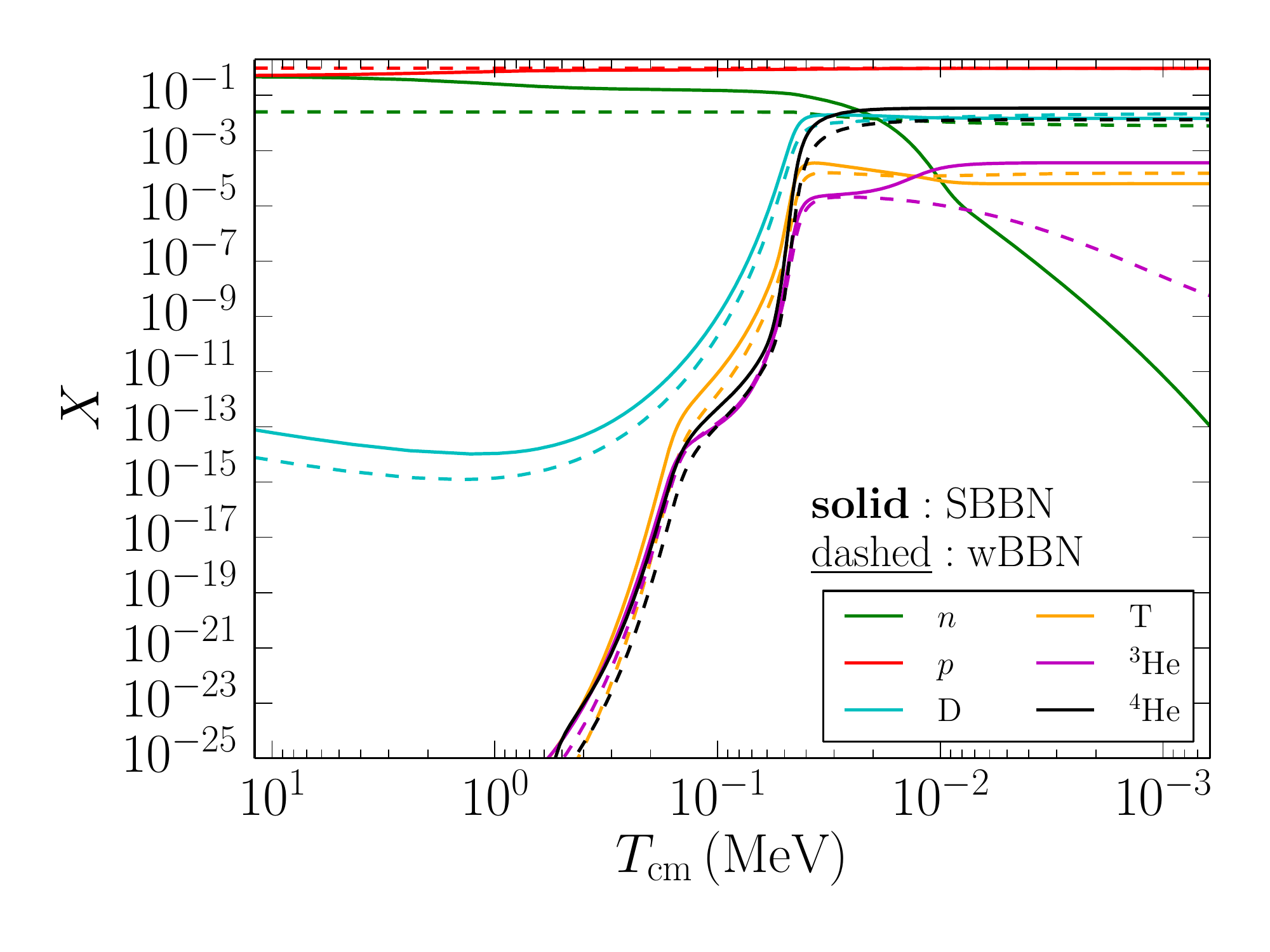}
  \end{center}
  \caption{\label{fig:comp_s2} Mass fractions, $X$, as a function of \tcm.
  Each color represents a different isotope.
  (Top) Solid lines are a SBBN calculation, and dashed lines are a weakless
  calculation where  $\np=1$.  The dashed green line for $n$ is coincident with
  the dashed red line for $p$.  $\eta=4.0\times10^{-12}$ for both calculations.
  (Bottom) Same as the top panel except $\np=0.026$ for the wBBN calculation.
  }
\end{figure}

Figures \ref{fig:comp_s1} and \ref{fig:comp_s2} show mass fractions versus \tcm
in a number of BBN cases.  Solid lines are from a SBBN calculation with an
evolving \np ratio and dashed lines from a wBBN calculation with a constant \np
ratio.  Both calculations, for a given figure, use the same value of $\eta$.
For ease in reading, we only plot the mass fractions of free neutrons ($n$),
free protons ($p$), deuterium (${\rm D}$), tritium (${\rm T}$), helium-3
(\hethree), and helium-4 (\heiv).  We note that the mass fractions of \livi,
\livii, and \bevii are all subsidiary to that of the hydrogen and helium
isotopes in both SBBN and wBBN.  We do not include other possibly stable
isotopes in a weakless universe, i.e., $^6{\rm He}$.

Figure \ref{fig:comp_s1} uses $\eta=6.075\times10^{-10}$ which is consistent
with the value in our universe \cite{PlanckXIII:2015}.  $\np=1$ for the wBBN
calculation in the top panel.  At high temperatures, the abundances are in
Nuclear Statistical Equilibrium (NSE).  NSE abundances are functions of
plasma temperature, $\eta$, $X_n$, $X_p$, and nuclear properties \cite{B2FH}.  The
mass fractions of ${\rm D}$, ${\rm T}$, and \hethree are all following NSE tracks above
$T\gtrsim 1\,{\rm MeV}$ (\heiv also follows a NSE track below the scale of the
vertical axis).  For temperatures $T>\deltamnp$ in a SBBN calculation, Eq.\
\eqref{eq:np_evolution} shows that $\np\sim 1$, implying that the SBBN and wBBN
scenarios have the same conditions for NSE.  As a result, the mass fractions
evolve identically for both SBBN and wBBN at high temperatures in the top
panel.  Although the mass fractions remain in NSE, the SBBN curves begin to
diverge from the wBBN curves once $T$ becomes comparable to \deltamnp.  At even
lower temperatures, \np for SBBN is well below unity and the mass fractions
diverge for the two scenarios.   Note that the red dashed curve ($X_p$ in
wBBN) is on top of the green dashed curve ($X_n$ in wBBN) in the top panel in Fig.\
\ref{fig:comp_s1}.  The wBBN mass fractions for $n$, $p$, and ${\rm D}$ have
frozen out at the end of the horizontal plotting axis.  Although neutrons and
protons can interact to form deuterons, the baryon density is so low that all
three mass fractions have frozen out at roughly the $1\%$ level.

For SBBN at $\eta=6.075\times10^{-10}$, \heiv freezes-out with a mass fraction
$X_{\heiv}\equiv Y_P\simeq0.25$, where we adopt the cosmological notation of
$Y_P$ to denote the mass fraction of \heiv.  The top panel of Fig.\
\ref{fig:comp_s1} shows that the number of baryons in ${\rm D}$ and \hethree is
only a few in $10^5$, implying that the vast majority of neutrons are in \heiv
nuclei.  Therefore, $\np\simeq1/7=0.14$ at the conclusion of SBBN.  The bottom
panel of Fig.\ \ref{fig:comp_s1} shows the same SBBN calculation as the top
panel, but the wBBN calculation now has $\np=0.14$.  In the NSE regime at high
temperatures, there is a clear difference between the SBBN and wBBN
calculations.  As the temperature decreases and the \np ratios converge to one
another for the two scenarios, the abundances begin to come into agreement,
especially for $p$ and \heiv.

Nuclear reactions which involve the capture of a neutron are not subject to the
Coulomb repulsion unavoidable in proton capture reactions.  At low
temperatures, neutrons can continue to capture on heavier nuclei or free
protons.  It is possible for BBN to occur over a long period if free neutrons
are present and the proper baryon number density is large.  For the wBBN
calculation in the top panel of Fig.\ \ref{fig:comp_s1}, there is a
preponderance of free neutrons at $\tcm\simeq10\,{\rm keV}$, although the
number density of baryons is low enough that there is no late-time rise in
${\rm D}$.  However, the flux of neutrons is large compared to the number of
\hethree targets.  The principal reactions which utilize neutrons and \hethree
nuclei as reactants, namely $\hethree(n,p){\rm T}$ and $\hethree(n,\gamma)\heiv$,
have not frozen-out at the end of the plotting axis.  As a result, the \hethree
mass fraction continues to decrease at the end of the simulation.  Similarly,
the flux of neutrons compared to \livi, \livii, and \bevii targets is also
large in the weakless scenario.  If we had plotted the mass fractions of the Li
and Be isotopes, they would also be decreasing at the end of the horizontal
plotting axis in much the same manner as the mass fraction for \hethree does.
We verified our hypotheses by extending the wBBN calculation down to
temperature of $T=100\,{\rm eV}$.  However, our library for the integrated
cross sections (based off of that in Ref.\ \cite{SMK:1993bb}) is not accurate
at such low temperatures.  Furthermore, the error in not finishing the wBBN
calculation would be small as the mass fractions of \hethree and the heavier
isotopes are orders of magnitude smaller than the more abundant species.  As we
consider the composition of the later weakless universe, we will take the mass
fractions from wBBN at the $T\sim 1\,{\rm keV}$ epoch and ignore the small
error present in \hethree.

In Fig. \ref{fig:comp_s2}, we use a value of $\eta=4.0\times10^{-12}$.  This
specific value is in line with the value chosen in Ref.\
\cite{HKP:2006_weakless}.  The top panel of Fig.\ \ref{fig:comp_s2} shows the
evolution of the mass fractions for SBBN and wBBN with $\np=1$.  For SBBN, the
mass fractions of ${\rm D}$, ${\rm T}$, and \hethree are all increased over the
freeze-out mass fractions when $\eta=6.075\times10^{-10}$ in Fig.\
\ref{fig:comp_s1}.  \heiv is significantly reduced for the low value of $\eta$
in Fig.\ \ref{fig:comp_s2}.  Conversely, in wBBN there are roughly equal
amounts of $n$, $p$, ${\rm D}$, and \heiv.  The freeze-out mass fractions are
in close agreement with Fig.\ [1] in Ref.\ \cite{HKP:2006_weakless}.  As
compared to the top panel of Fig.\ \ref{fig:comp_s1}, neutrons are not
predominantly incorporated into \heiv nuclei for the lower value of $\eta$.  In
both SBBN and wBBN, the ${\rm D}$ mass fraction is comparable to that of \heiv.
SBBN makes less \heiv (and less ${\rm D}$) as the \np ratio evolves to 0.026.

The bottom panel of Fig.\ \ref{fig:comp_s2} is identical to the top panel
except we run wBBN with $\np=0.026$.  Although the \np ratios are identical
between SBBN and wBBN, the final freeze-out mass fractions of ${\rm D}$ and
\heiv differ.  Furthermore, in the weakless scenario, there is a larger mass
fraction of ${\rm D}$ as compared to \heiv.  With fewer neutrons, the NSE
abundances are lower.  We observe this by using the top and bottom panels of
Fig.\ \ref{fig:comp_s2} to compare the locations of the dashed lines of ${\rm
T}$ and \heiv with respect to the solid SBBN lines.  The initial conditions for
out-of-equilibrium nucleosynthesis occur when the \np ratio is still evolving
in SBBN, i.e., $\np>0.026$ when the mass fractions go out of equilibrium.  As a
result, the sum of $X_{\rm D}$ and $Y_P$ is larger in SBBN than wBBN.  For the
\np ratios to be equal at freeze-out, the deficit of neutrons in wBBN must be
incorporated into a different nuclide.  Indeed, the bottom panel of Fig.\
\ref{fig:comp_s2} shows a mass fraction of free neutrons on order of $1\%$.
This is a key difference of SBBN compared to wBBN which was absent in Fig.\
\ref{fig:comp_s1}: identical \np ratios at freeze-out do not imply identical
mass fractions of ${\rm D}$ and \heiv.  If the baryon number is low enough,
out-of-equilibrium nucleosynthesis begins during weak freeze-out.  The \np
ratio alone is not enough to predict the final mass fractions of \heiv.

To conclude this section, we note neither SBBN nor wBBN possesses ${\rm D}$ and
${\rm T}$ peaks in Fig.\ \ref{fig:comp_s2}.  The peaks are quite visible in the
bottom panel of Fig.\ \ref{fig:comp_s1} at $\tcm\sim50\,{\rm keV}$ for both
SBBN and wBBN.  The peaks stem from synthesis of ${\rm D}$ and ${\rm T}$ into
larger nuclei, most notably \heiv, and no production channels of equivalent
strength.  For the lower value of $\eta$, the nuclear reaction rates freeze-out
at an earlier time and the mass fractions plateau.

\subsection{Parameter space scan}\label{ssec:contour}

The previous section detailed a comparison between SBBN and wBBN at various
values of $\eta$.  In this subsection, we explore the \np versus $\eta$
parameter space of wBBN.  We cannot directly compare with SBBN because there is
no dual parameter space for cosmological inputs [see Fig.\ (2) in Ref.\
\cite{AAGM:2017_vac} for SBBN mass fractions as a function of $\eta$].  As an
alternative, we place a red star on the contour plots at the values
$\eta=6.075\times10^{-10}$ and $\np=0.14$ of our universe.  The red star comes
from a simulation that models our universe with the weak interaction, namely
the SBBN calculation in Fig.\ \ref{fig:comp_s1}.  \np changes with time in such
a scenario.  In addition, our universe contains neutrino energy density so the
Hubble expansion rate is different.  We emphasize that the red star is only for
illustrative purposes and does not belong to the class of weakless universes.
It is only meant as a guide and not for direct comparison.

\begin{figure}
  \begin{center}
    \includegraphics[width=\columnwidth]{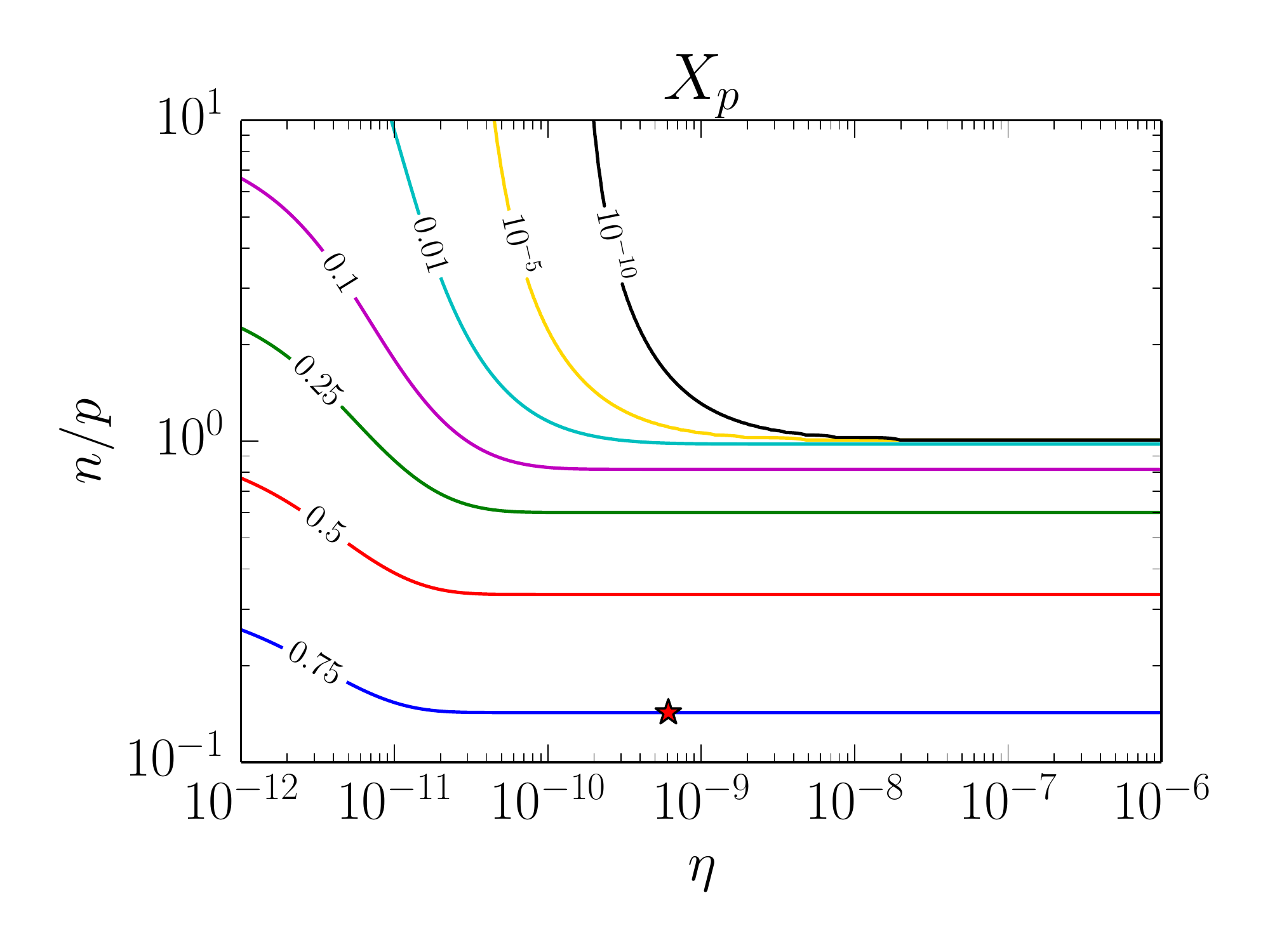}
  \end{center}
  \caption{\label{fig:h1} The neutron-to-proton ratio, \np, versus the
  baryon-to-photon ratio, $\eta$, at contours of constant $p$ mass fraction.  In
  a weakless universe, \np is fixed at the start of BBN and does not evolve
  through nuclear freeze-out.  In our universe, weak interactions change \np
  until \heiv formation when $\np\approx1/7$.  The red star indicates the point
  in parameter space where \np terminates its evolution in our universe.  This
  point should only be taken for illustrative purposes and not labeled a member of the
  weakless class of universes.
  }
\end{figure}

\begin{figure}
  \begin{center}
    \includegraphics[width=\columnwidth]{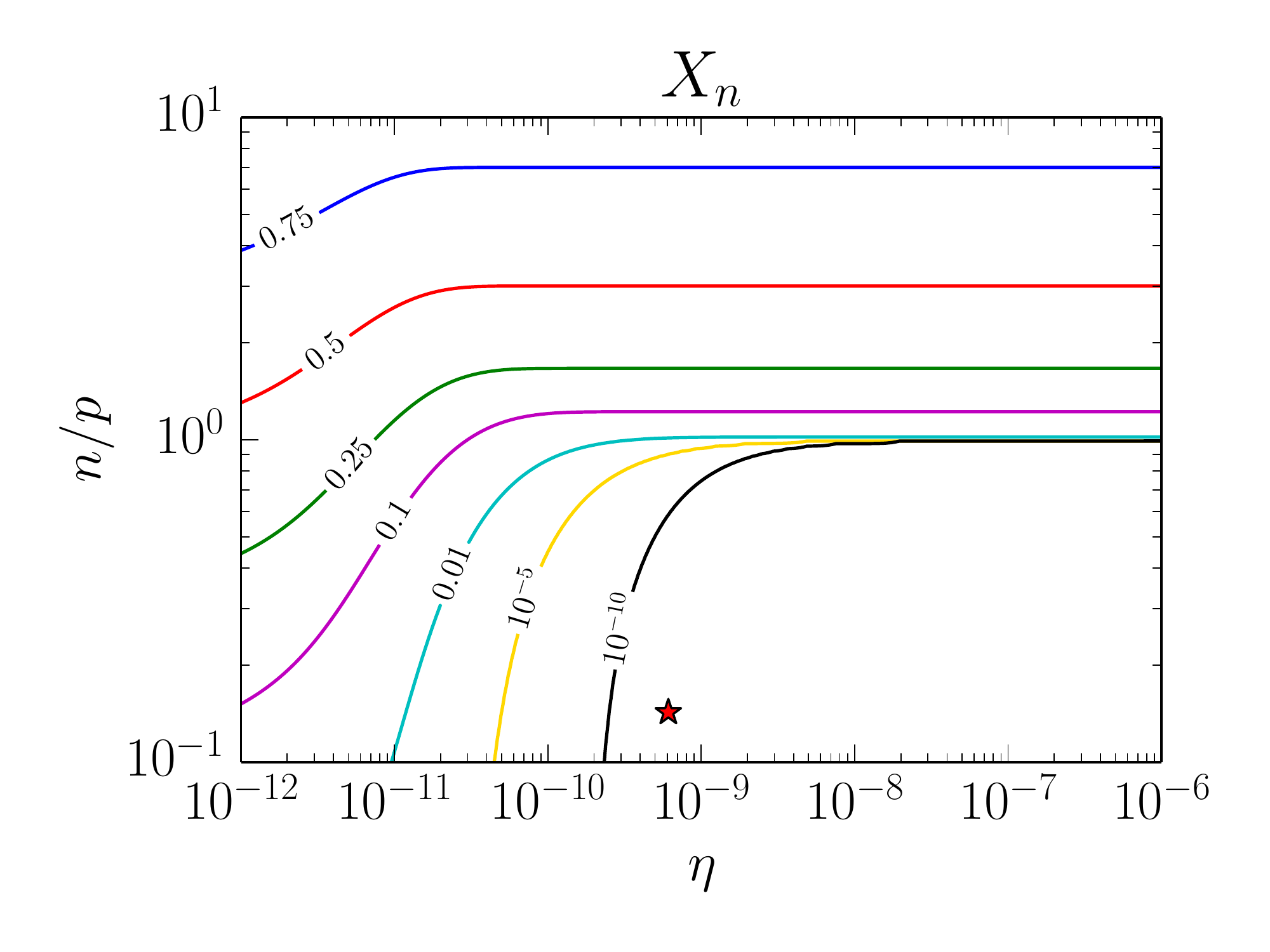}
  \end{center}
  \caption{\label{fig:neut} \np versus $\eta$ at contours of constant $n$ mass
  fraction.  The red star indicates the point in parameter space where \np
  terminates its evolution in our universe.}
\end{figure}

Figure \ref{fig:h1} shows contours of constant $p$ mass fraction in the $\np\,$
-- $\eta$ plane.  The red star is located close to the $75\%$ contour, which is
in agreement with the value from SBBN as seen on the bottom panel of Fig.\
\ref{fig:comp_s1}.  For the range of $\eta$ plotted, a significant fraction of
the baryons are free protons if \np is below unity.  Once \np becomes larger
than unity, few free protons remain.  The isospin mirror of Fig.\ \ref{fig:h1}
is Fig.\ \ref{fig:neut}: contours of constant $X_n$ in the $\np\,$ -- $\eta$
plane.  For \np above unity, there exists a significant fraction of free
neutrons.  Conversely, for \np below unity, few free neutrons remain.

\begin{figure}
  \begin{center}
    \includegraphics[width=\columnwidth]{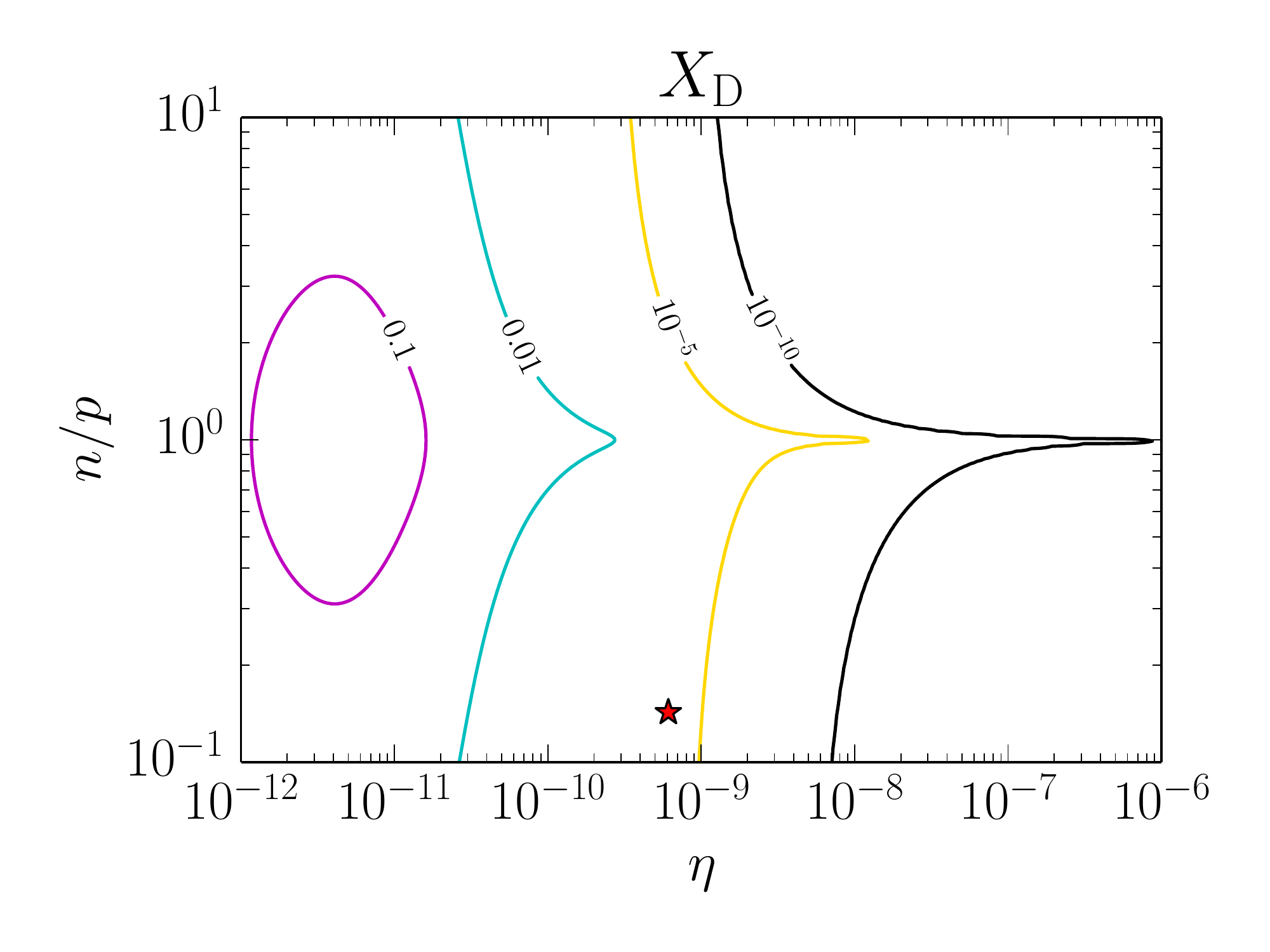}
  \end{center}
  \caption{\label{fig:h2}  \np versus $\eta$ at contours of constant ${\rm D}$ mass
  fraction.  The red star indicates the point in parameter space where \np
  terminates its evolution in our universe.}
\end{figure}

Figures \ref{fig:h1} and \ref{fig:neut} show a remarkable degree of symmetry
about the line $\np=1$.  The symmetry is broken, albeit slightly, in the mass
fraction of ${\rm D}$.  Figure \ref{fig:h2} shows contours of constant $X_{\rm D}$ in
the $\np\,$ -- $\eta$ plane.  At constant $\eta$, the mass fraction of ${\rm D}$
increase as \np approaches unity from either direction.  The rate of increase
is larger as \np decreases towards unity.  This is only evident for large
$\eta$ in Fig.\ \ref{fig:h2}.  The contours of $X_{\rm D}$ at $10^{-5}$ and
$10^{-10}$ are closer in the parameter space for $\np>1$.  The value of the
contours indicates that the degree of asymmetry is small -- a few parts in
$10^{5}$.

The red star in Fig.\ \ref{fig:h2} is located on the $X_{\rm
D}=2.7\times10^{-5}$ contour.  For comparison, the ${\rm D}$ mass fraction of
SBBN is $4.0\times10^{-5}$.  wBBN is able to produce a large mass fraction of
${\rm D}$.  The largest mass fraction of ${\rm D}$ is $X_{\rm D}=0.14$ at
$\np=1$ and $\eta=5\times10^{-12}$.

\begin{figure}
  \begin{center}
    \includegraphics[width=\columnwidth]{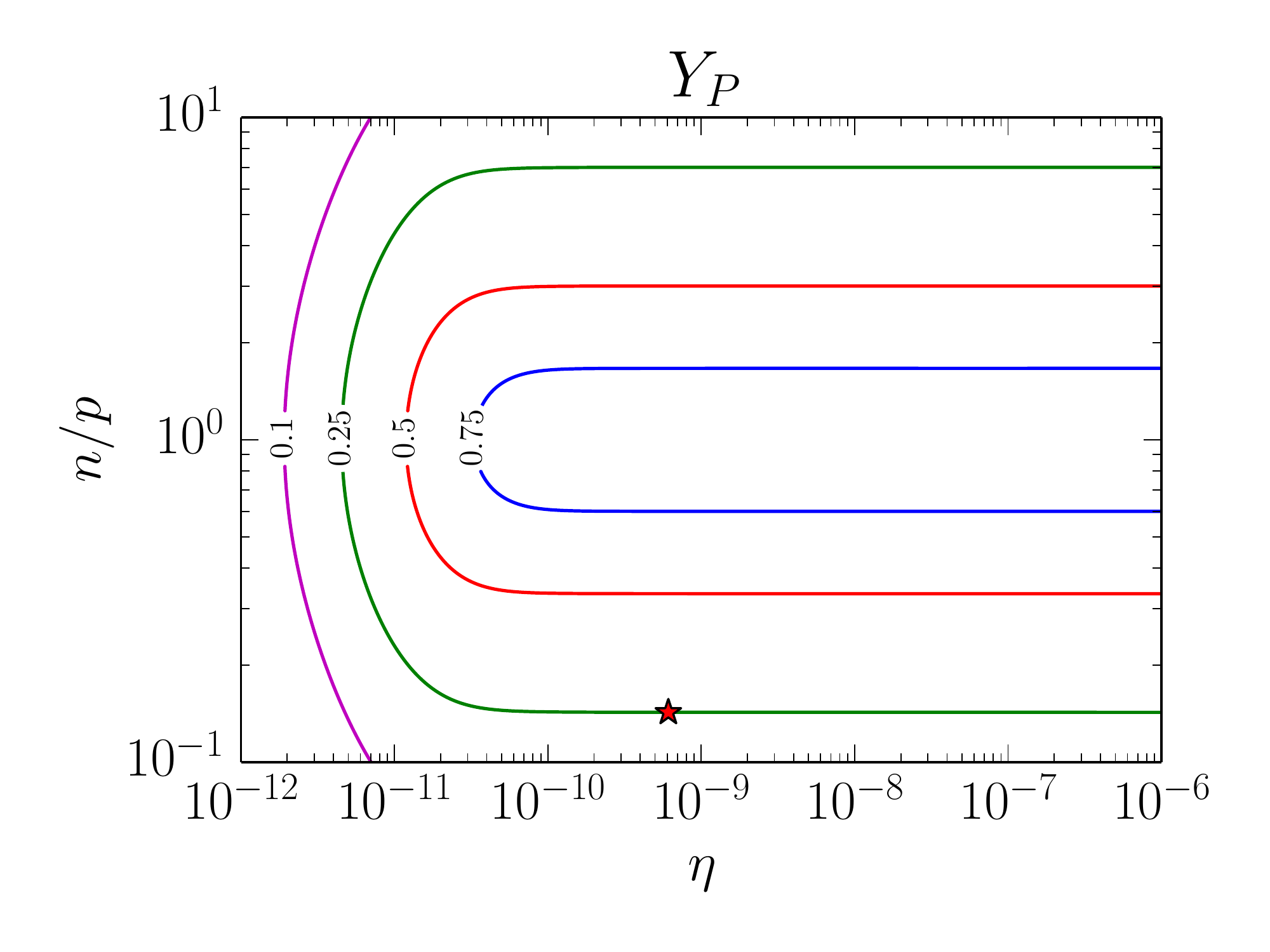}
  \end{center}
  \caption{\label{fig:yp} \np versus $\eta$ at contours of constant \yp, the
  \heiv mass fraction.  The red star indicates the point in parameter space where
  \np terminates its evolution in our universe.}
\end{figure}

The symmetry about $\np=1$ is restored in Fig.\ \ref{fig:yp}: contours of
constant $Y_P$ in the $\np\,$ -- $\eta$ plane.  For the parameter space
shown in Fig.\ \ref{fig:yp}, the smallest value of $Y_P$ is
$\sim7\times10^{-3}$.  This value occurs for small $\eta$ and either large or
small \np.  The mass fraction value is above our asymmetry limit of $10^{-5}$,
so we cannot detect any visible asymmetry in Fig.\ \ref{fig:yp}.  For small
$\eta$, Fig.\ \ref{fig:yp} shows that not all neutrons are incorporated into
\heiv nuclei.  In fact, the mass fraction of ${\rm D}$ is larger than that of \heiv
for a certain range of $\eta$ if $\np=1$.  Furthermore, Fig.\ \ref{fig:neut}
shows that most neutrons are free at small $\eta$.  For large $\eta$, the mass
fraction of \heiv is independent of $\eta$ in the range of \np we employ in Fig.\
\ref{fig:yp}.  The contours of constant mass fraction are symmetrical about
$\np=1$ and horizontal.  We can succinctly capture the relationship between the
mass fraction of \heiv and \np in this range
\beq\label{eq:yp_np}
  \yp \simeq 2\,\frac{\min(1,\np)}{1+\np}.
\eeq
If $\np<1$, Eq.\ \eqref{eq:yp_np} reduces to the familiar $\yp=2(\np)/(1+\np)$
\cite{1990eaun.book.....K}.  Finally, we note that the location of the contours
in Fig.\ \ref{fig:yp} (or, more specifically, the rounded ends at low $\eta$)
depends on the nuclear reaction rates versus the Hubble expansion rate.  We do
not include neutrinos or any other form of dark radiation in our model of
wBBN.  If we had, that would increase the Hubble rate and decrease $Y_P$.
The result would be a shift of the contours in Fig.\ \ref{fig:yp} in the
horizontal direction towards higher $\eta$.  There would be no change in the
vertical direction of Fig.\ \ref{fig:yp} at the level of precision presented in
the parameter space.

Table \ref{tab:yields} gives the freeze-out mass fractions at numerous values
of $\eta$ which we will use in sections \ref{sec:galaxy} -- \ref{sec:stars}.
$\np=1$ for all values of $\eta$.  It appears that the ${\rm D}$ mass fraction gets
closer to the $n$ and $p$ mass fractions for increasing $\eta$.  For $\np=1$,
${\rm D}$ gets closest to $n$ when $\eta\simeq10^{-7}$ at a value
$X_{\rm D}/X_n=0.95$.  ${\rm D}$ gets closest to $p$ when $\eta\simeq5\times10^{-8}$
at a value $X_{\rm D}/X_{p}=0.86$.  The \hethree mass fraction is much lower
than the mass fractions for the other nuclides, as explained in Figs.\
\ref{fig:comp_s1} and \ref{fig:comp_s2} by the lack of a Coulomb barrier in
neutron capture.  By the same logic, Table \ref{tab:yields} shows a slight
excess of the $p$ mass fraction over that of free neutrons even though the universe
is isospin symmetric, i.e., $\np=1$.  This is due to the high reactivity of
neutrons, or equivalently, a Coulomb barrier in proton capture.

\begin{table*}
  \begin{center}
  \begin{tabular}{c c c c c}
    \hline
    $\eta$ & & $10^{-11}$ & $10^{-10}$ & $10^{-9}$\\
    \midrule[1.0pt]
    $n$ & & $0.2136$ & $3.420\times10^{-2}$ & $3.535\times10^{-3}$ \\
    $p$ & & $0.2139$ & $3.428\times10^{-2}$ & $3.545\times10^{-3}$ \\
    ${\rm D}$ & & $0.1197$ & $2.615\times10^{-2}$ & $2.766\times10^{-3}$ \\
    ${\rm T}$ & & $9.764\times10^{-4}$ & $2.320\times10^{-4}$ & $2.469\times10^{-5}$ \\
    \Heiii & & $8.844\times10^{-10}$ & $2.605\times10^{-10}$ & $2.819\times10^{-11}$ \\
    \heiv & & $0.4518$ & $0.9051$ & $0.9901$ \\
    \hline
  \end{tabular}
  \end{center}
  \caption{\label{tab:yields}Mass fractions at different values of $\eta$.  All
  values for $\np=1$.
  }
\end{table*}

\section{Galaxy Formation}\label{sec:galaxy} 

This section considers how nuclear reactions involving free neutrons can
potentially affect the galaxy formation process. Here we assume that structure
formation involves processes that are analogous to those acting in our
universe. Without weak interactions, a universe can in principle still produce
dark matter \cite{HKP:2006_weakless}. In this case, the timing of structure
formation would remain the same, and this is the case considered here. 

For completeness we note that even in the absence of dark matter, a purely
baryonic universe can produce structure. In the present context, however, the
baryon-to-photon ratio $\eta$ must be smaller due to BBN constraints (see Sec.\
\ref{ssec:contour}) so that the epoch of matter domination occurs at lower
redshift.  As a result, galaxy formation takes place at later epochs when the
universe is more diffuse, so that galaxies are less dense for a fixed value of
the amplitude $Q$ of density fluctuations. For the particular value $Q=10^{-5}$
realized in our universe, the resulting densities of galaxies could be so low
that baryons have difficulty cooling and condensing \cite{tegrees,tegmark}.
This issue can be alleviated with larger values of the fluctuation amplitude
$Q$ \cite{tegmark,coppess}.  With no dark matter (which has $\sim6$ times the
density of baryons in our universe) and lower $\eta$ (by a factor of $\sim50$),
the total matter density is smaller by a factor of $\sim300$. The corresponding
value of the fluctuation amplitude $Q\sim3\times10^{-3}$.

\subsection{Properties of Dark Matter Halos}\label{sec:properties} 

Simulations of structure formation show that dark matter halos asymptotically
approach a nearly universal form. The density profile of the halo is thus taken
to be a Hernquist profile of the form  
\be
\rho(r) = {\rho_0 \over \xi (1+\xi)^3} \,,
\label{hqprofile} 
\ee
where the dimensionless coordinate $\xi$ is defined via
\be
\xi = {r \over r_0} \,.
\ee
Note that we use a slightly steeper power-law for the density distribution
($\rho\sim\xi^{-4}$ for $\xi\gg1$) compared to the often-used NFW profile
\cite{nfw} ($\rho\sim\xi^{-3}$). Although only the inner part of the potential
well matters for the considerations of this paper, we use the form
(\ref{hqprofile}) because it has a finite mass and because numerical
simulations show that halo density profiles become steeper at later epochs
\cite{busha2005,busha2007}. Given the form of equation (\ref{hqprofile}), the
corresponding potential and enclosed mass have the forms 
\be
\Psi = {\Psi_0 \over 1 + \xi} 
\qquad {\rm and} \qquad 
M(r) = M_T \left( {\xi \over 1 + \xi} \right)^2\,.
\ee
We can take $\rho_0$ and $r_0$ to be the defining parameters of the halo. The
scale $\Psi_0$ for the potential and the total mass $M_T$ are then given by
\be
\Psi_0 = 2 \pi G \rho_0 r_0^2 
\qquad {\rm and} \qquad 
M_T = 2\pi \rho_0 r_0^3 \,. 
\ee

For the example of the Milky Way galaxy, we can model the halo with a profile
of the form given by equation (\ref{hqprofile}) if we take the scale length
$r_0$ = 65 kpc and the density scale $\rho_0$ = $10^{-25}$ g cm$^{-3}$ (see
Ref. \cite{spirograph} and references therein).

\subsection{Hydrostatic Equilibrium}\label{sec:hydrostatic}

In order to assess the possible effects of nuclear reactions on galaxy
formation, we need to determine the temperature and density
distributions of the baryonic gas. We start with an order of magnitude
estimate: the usual assumption is that dark matter collapses first, and
then gas falls into the dark matter halo. The collapse of the gaseous
component and subsequent shocks heat up the material to a temperature
given by 
\be
3kT = {G M \mpro \over r} \,,
\ee
where $M=M(r)$ is the enclosed mass of the dark matter halo at radius $r$. For
the galaxy profiles considered here, this expression becomes 
\be
3kT = {G M_T \mpro \over r_0} {\xi \over (1+\xi)^2}\,.
\ee
For the Milky Way values, the benchmark 
temperature scale is given by 
\be
T = {G M_T \mpro \over 3k r_0} \approx 2.7 \times 10^6 {\rm K}\,.
\label{tempguess} 
\ee
This estimate is approximate and does not require that the 
gaseous density and temperature profiles approach an 
equilibrium condition. We thus generalize the treatment
in the following discussion. 

For a given dark matter halo, we can determine the density 
and temperature profile of the gas under the assumptions that:
(1) the dark matter halo dominates the gravity of the system; 
(2) the gas can be considered in hydrostatic equilibrium; and 
(3) the equation of state for the gas is polytropic 
\be
P = K \rho^\Gamma\equiv K \rho^{1 + 1/n} \,,
\ee
where $n$ is the usual polytropic index and $K$ is a scaling constant.
Hydrostatic equilibrium thus implies 
\be
K \Gamma \rho^{\Gamma-2} {d \rho \over d\xi} = 
- {\Psi_0 \over (1+\xi)^2} \,, 
\ee
so that the density profile has the form 
\be
\rho(\xi) = 
\left[ {(\Gamma-1)\Psi_0 \over K\Gamma} \right]^{1/(\Gamma-1)} 
(1 + \xi)^{-1/(\Gamma-1)} \,. 
\ee
To fix ideas, consider the standard case of an adiabatic equation of state for
a monoatomic gas, where $\Gamma=5/3$.  The density profile
then becomes 
\be
\rho(\xi) = 
\left[ {2\Psi_0 \over 5K} \right]^{3/2}
(1 + \xi)^{-3/2} \equiv \rho_X (1 + \xi)^{-3/2} \,, 
\ee
where the second equality defines a benchmark density scale. The total mass in
baryons is given by the integral 
\be\label{eq:mb1}
M_b = 4\pi r_0^3 \rho_X \int_0^{\xi_{\rm max}} 
{\xi^2 \over (1+\xi)^{3/2}} d\xi 
\equiv 4\pi r_0^3 \rho_X I(\xi_{\rm max}) \,,
\ee
where the final equality defines the dimensionless integral $I$.  Note that we
must invoke a finite boundary to the density distribution of the baryons in
order to keep the mass finite (for this profile).  If we define $f_b$ to be the
baryonic fraction of the total mass, then
\be\label{eq:mb2}
M_b = f_b M_T = f_b 2\pi \rho_0 r_0^3 \,.
\ee
Equating Eqs.\ \eqref{eq:mb1} and \eqref{eq:mb2} implies
\be
2\rho_X I = f_b \rho_0 \qquad {\rm or} \qquad 
\rho_X = {f_b \rho_0 \over 2I}\,.
\ee
The integral $I$ is of order unity. For example, if $\xi_{\rm max}=1$, the integral
$I\sim1/5$. In general, we want to consider somewhat larger values of
$\xi_{\rm max}$, so we can take $2I\approx1$ and hence use the ansatz 
\be
\rho_X = f_b \rho_0 \,, 
\label{rhox} 
\ee
where $f_b\sim1/6$ in our universe.  Next we note that since 
\be
P = K \rho^\Gamma = \rho {kT \over \mpro} \,,  
\ee
where \mpro is the proton mass, we find that 
\be
kT = \mpro K \rho^{\Gamma-1} = \mpro K \rho^{2/3} 
= \mpro K \rho_X^{2/3} (1+\xi)^{-1} \,. 
\ee
We can thus write
\be
T = {T_X \over 1 + \xi} \qquad {\rm where} \qquad 
T_X \equiv {\mpro K \over k} \rho_X^{2/3} = 
{2\mpro\over5k} \Psi_0\,.
\ee
The temperature in a potential well is typically assumed to be of order
$kT\sim\mpro\Psi$. For the particular case of an $n=2/3$ polytrope considered
here, the last equality expresses the particular realization of this expected
relation. For halos with properties of the Milky Way, the temperature scale
$T_X \approx 8 \times 10^6$ K, which is roughly comparable to the original
estimate from equation \eqref{tempguess}. 

\subsection{Column Density} 
\label{sec:column}

The discussion thus far has assumed that the radiation produced by any possible
nuclear reactions escapes the halo and does not affect its structure. To verify
the validity of this assumption, we need to estimate the optical depth of the
halo of its internal radiation. The first step is to determine the column
density.  To wit, the column density of the dark matter in the halo, integrated
from spatial infinity to a radial location $\xi$, is given by the expression 
\be
N_{\rm dm} (\xi) = \rho_0 r_0 \left\{
\log\left[ {1 + \xi \over \xi} \right] - 
{2\xi + 3 \over 2(\xi+1)^2} \right\} \, . 
\ee
We can use the hydrostatic profiles from the previous section to obtain the
column density of the gaseous component. For simplicity, we use the solutions
for $\Gamma=5/3$, and find a column density in gas
\be
N_{\rm gas} (\xi) = 2\rho_X r_0 (1 + \xi)^{-1/2} \approx 
2 f_b \rho_0 r_0 (1 + \xi)^{-1/2} \,,
\ee 
where we have used Eq.\ \eqref{rhox}.  The benchmark value
$\tau_0$ of the optical depth of the halo to its radiation field --
that generated by nuclear reactions --- can be defined as
\be
\tau_0 = 2 f_b \rho_0 r_0 {\sigma_T \over \mpro} \,,
\ee
where we assume that the cross section for interactions between the
gamma rays from the nuclear reactions and the remaining gas is given
by the Thomson cross section $\sigma_T$.  For values in our universe,
this optical depth $\tau \sim 10^{-3}$.  As a result, the halo does
not have a photosphere -- it is optically thin to the radiation it
generates, so that photons freely stream outwards. In the inner
regions of the galaxy, the optical depth approaches $\tau_0$. In the
outer regions, the the optical depth has spatial dependence given by
\be
\tau (\xi) \approx \tau_0 (1 + \xi)^{-1/2}\,.
\label{tauvsxi} 
\ee

\subsection{Heating and Cooling Rates}\label{sec:cooling}

The cooling rate per unit volume can be generically written in the form 
\be
{dE \over dt dV}\biggr|_{\rm cool} = n_e n_p \Lambda(T) = 
n_e n_p \langle \sigma v \rangle_{\rm cool}
\epcool \,,
\label{cooling} 
\ee
where $n_e$ and $n_p$ are the number densities of electrons and free protons,
respectively.  For sufficiently high gas temperature, the cooling process is
primarily due to bremsstrahlung scattering \cite{carr}, so the cross section is
close to the Thomson cross section $\sigma_T$ and the speed is given by the
thermal speed of the electrons $v_s=(kT/m_e)^{1/2}$.  The energy lost per
scattering $\epcool$ can be written in the form 
\be
\epcool = {4e^2\over\lambda} \approx 2.37\,\,{\rm keV}
\qquad {\rm where} \qquad \lambda = {h \over m_e c} \,.
\ee
Note that these expressions are approximate -- an accurate treatment requires
integration of the interactions over the thermal distribution of particles and
results in corrections given by factors of order unity. We can thus write the
cooling rate per unit volume in the form 
\be
{dE \over dt dV}\biggr|_{\rm cool} = A n_e n_p \sigma_T 
\left({kT \over m_e} \right)^{1/2} {4 e^2 m_e c \over h} \,,
\label{heating} 
\ee
where $A$ is a dimensionless parameter of order unity.  Although Eq.\
\eqref{heating} is highly simplified, the resulting cooling times are
comparable to those found previously \cite{whiterees,galli}.

This cooling treatment assumes fully ionized gas and hence high temperatures.
As the temperature falls to about $T \sim 10^4$ K, this process becomes
ineffective, and the cooling rate becomes much smaller. As a result, halo gas
tends to cool down to $10^4$ K and then stay at that temperature.

The corresponding heating rate due to nuclear reactions can be written in the
form 
\be
{dE \over dtdV}\biggr|_{\rm heat} = n_p n_n \sigmanp 
\epnuke \,,
\ee
where $\sigmanp\sim10^{-19}\,{\rm cm}^3\,{\rm s}^{-1}$ and $\epnuke$ is the
energy deposited in the gas due to the reaction. 

Most of the energy from the reaction is contained in gamma rays, which are
(mostly) optically thin and hence tend to leave the system. The deuterium
nucleus itself experiences a recoil energy of about 1 keV, so that the
deposited energy has a lower bound $\epnuke > 1$ keV.

We can find the requirements for the heating rate \eqref{heating} and the
cooling rate \eqref{cooling} to be in balance:
\begin{align}
&A \sigma_T \left({kT \over m_e} \right)^{1/2} \epcool
= \chi_n \sigmanp \epnuke\nonumber\\
\implies&T = {m_e\over k} \left( {\sigmanp \epnuke \over 
A \sigma_T \epcool} \right)^2 \chi_n^2 \,.
\label{tempcoolnuke} 
\end{align}
The parameter $\chi_n=n_n/n_p$ is the relative abundance of free neutrons with
respect to free protons [and not equivalent to \np from Eq.\
\eqref{eq:np_def}]. Note that we have also assumed that the gas is fully
ionized so that $n_e=n_p$. 

For optically thin conditions, where the gamma rays from the nuclear reactions
escape, the two energy scales are comparable ($\epnuke\approx\epcool$) and the
equilibrium temperature would be $T<1$ K. The approximations used in the
cooling function break down well before this temperature is reached, so that we
expect the gas to stay at $T\sim10^4$ K. In the other limit, where all of the
energy from the nuclear reactions is contained within the gas, then $\epnuke
\sim 1$ MeV and the equilibrium temperature would be $T\sim8\times10^4$ K.

As a result, under optically thin conditions, the heating due to nuclear
reactions is ineffective and the heating and cooling of gas on galactic scales
proceeds in the usual fashion. The heating due to nuclear processes would only
become important if the temperature from Eq.\ \eqref{tempcoolnuke} exceeds
$T\sim10^4$ K. This requirement, in turn, implies that the energy deposited per
reaction $\epnuke\simgreat1$ MeV. In order for this much energy to be retained
in the gas, the optical depth must be close to unity, but Eq.\ \eqref{tauvsxi}
implies that $\tau$ reaches a maximum value of $\tau_0\ll1$.  As a result, the
galaxy remains optically thin to the radiation generated by nuclear reactions.

\subsection{Time Scales}\label{sec:timescales}

The collapse time is given by 
\be
t_G = (G\rho)^{-1/2} \approx 95\, {\rm Myr} 
\left( {n \over 1\, {\rm cm}^{-3}} \right)^{-1/2} 
\label{tcollapse}
\ee
The cooling time is given by 
\begin{align}
t_{\rm cool} = {3kT \over 2 n \Lambda} &= 
{3kT \over 2An \sigma_T v_s \epcool}\nonumber\\
&\approx 5\, {\rm Myr} \left( {n \over 1\, {\rm cm}^{-3}} \right)^{-1} 
\left( {T \over 10^6\, {\rm K}} \right)^{1/2} \,.
\label{tcool}
\end{align}
The heating time due to nuclear reactions is given by 
\begin{align}
t_{\rm nuke} &= {3kT \over 2n \sigmanp \epnuke}\nonumber\\
&\approx 41\, {\rm Gyr} \left( {n \over 1\, {\rm cm}^{-3}} \right)^{-1} 
\left({T \over\, 10^6 {\rm K}}\right) 
\left({\epnuke \over 1\,{\rm keV}}\right)^{-1} \,.
\label{tnuke}
\end{align}
The cooling time is much shorter than the heating time over the entire range of
applicability of the cooling function used here. Once the gas cools down to
$T\sim10^4$ K, the cooling processes become much less effective, and cooling
processes cease to operate. 

We can also find a benchmark density scale where the collapse time is equal to
the heating time. This scale is given by
\be
n_\star \approx 20 \,\,{\rm cm}^{-3} \,\,
\left({T \over 10^4\, {\rm K}}\right)^2 \,
\left({\epnuke \over 1\,{\rm keV}}\right)^2 \,.
\label{nbench} 
\ee
This number density is larger than the typical mean density of the interstellar
medium in the Galaxy, but smaller than the density of molecular clouds
($n\sim10^2-10^3$ cm$^{-3}$), where star formation takes place. 

In summary, nuclear reactions in this scenario do not affect structure
formation for scales larger than molecular clouds. Cloud formation and
subsequent evolution of star forming regions, however, can be affected and are
discussed in subsequent sections.

\subsection{Scaling with Amplitude of the Density Fluctuations} 
\label{sec:ampscaling}

Our universe has initial density fluctuations, inferred from the observed
inhomogeneities in the CMB radiation, with amplitude $Q \approx
10^{-5}$. In other universes, these fluctuations could be larger, with the
consequence that galaxies can form earlier.  This difference in timing, in
turn, results in galaxies that are denser. Here we define the relative
amplitude 
\be
q \equiv {Q \over Q_0} \,,
\ee
where $Q_0$ is the value in our universe. The parameters $\rho_0$ and
$r_0$ of the dark matter halos vary with the fluctuation amplitude
that specifies the initial conditions.  For dark matter halos with
density profiles given by Eq.\ \eqref{hqprofile}, the dependence
of $\rho_0$ and $r_0$ on the amplitude $Q$ has been derived previously 
\cite{coppess}, where these results are based on the standard paradigm
for galaxy formation \cite{whiterees}. The resulting scaling laws can
be written in the form
\be
\rho_0 \propto q^3 \qquad {\rm and} \qquad r_0 \propto q^{-1} \,. 
\label{rhoscale} 
\ee
The temperature scale $T_X$ is determined by the depth of the 
gravitational potential well of the halo. Since $\Psi_0\sim G\rho_0r_0^2$,
the temperature scales according to 
\be
T_X \propto q\,.
\label{tempscale} 
\ee
The fluctuation amplitude can be larger by a factor of $\sim1000$
\cite{coppess}. 

Now we consider how the time scales vary with changes in 
the fluctuation amplitude. Using Eqs.\ \eqref{rhoscale}
and (\ref{tempscale}), we find the scaling laws 
\be
t_G \propto q^{-3/2},\qquad 
t_{\rm cool} \propto q^{-5/2},\qquad {\rm and} \qquad 
t_{\rm nuke} \propto q^{-2}\,.
\ee

We first consider the case where the gas in the halo remains optically thin, so
that the deposited energy $\epnuke\sim1$ keV.  For the larger starting
temperature induced by larger $q$, the cooling rate dominates even more over
the heating rate due to nuclear reactions. The gas in these denser galaxies
will readily cool down to the temperature $T\sim10^4$ K where these cooling
processes become ineffective.  However, as the galaxies create their
substructures, they do so at higher densities, which can be larger than the
threshold where nuclear reactions play a role [see Eq.\ \eqref{nbench}].

Conversely, the optical depth scales as $\tau\propto\rho_0r_0$ $\propto q^2$.
For $q\simgreat30$, the halos become optically thick and retain most of the
energy generated by the nuclear reactions. The energy scale $\epnuke$ thus
increases from $\sim1$ keV to $\sim1$ MeV for sufficiently dense galaxies.

\section{Processes in the Interstellar Medium}\label{sec:ism}

\subsection{Molecular Clouds}\label{sec:clouds} 

The considerations of the previous section show that the additional heating due
to nuclear reactions does not greatly inhibit the cooling of gas on galactic
scales.\footnote[2]{We are considering universes with the same amplitude $Q$ of
the initial density fluctuations.}  In this section, we thus assume that
galaxies form in an analogous fashion to those in our universe and have the
same basic internal structures.  In the next level of structure formation, the
galaxy assembles molecular clouds, which in turn support the process of star
formation. Molecular clouds have densities of order $n\sim100$ cm$^{-3}$ on
their largest scales, with much denser internal structure. At these densities,
the nuclear reactions from stable free neutrons can act to prevent the cooling
of cloud material and hence delay star formation.

To start, we consider the simplest possible model of a molecular cloud: The
structure is assumed to have constant density with $n\approx100-300$ cm$^{-3}$.
Like clouds in our universe, the thermal pressure is much smaller than that
provided by both magnetic fields and turbulence. These latter sources of
pressure thus support the cloud, so that we can assume that its mechanical
structure is largely independent of the thermal evolution of the constituent
gas. 

The cooling processes for the gas are different from those of the previous
section. In this case, the gas is largely neutral (not ionized) and the cooling
processes become inefficient. In our universe, the cooling processes become
dominated by line emission from heavy elements, in spite of their low relative
abundances. In this context, however, we consider the gas to be composed only
of hydrogen, free neutrons, and helium. The cooling function will thus be
similar to that applicable for the formation of the first stellar generation in
our universe. In the limit of high density $n\to10^3$ cm$^{-3}$, the gas can
maintain local thermodynamic equilibrium (LTE) and the cooling function
$\Lambda \propto n_H$ \cite{galli}. Moreover, we can model the cooling function
with the from 
\be
{dE \over dt dV}\biggr|_{\rm cool}
= n_H C \left({T\over10^4\,{\rm K}}\right)^3\,,
\ee
where the constant 
\be
C =  10^{-19}\,{\rm erg}\,{\rm s}^{-1}\,,
\ee
and where the functional form is approximate. In the approximation of constant
cloud density, which holds in the absence of expansion or contraction of the
gas, the time evolution of the cloud material is governed by the equation 
\be
{3 \over 2} n {d \over dt} \left( kT \right) = 
- n C T_4^3 + n_p n_n \sigmanp \epnuke \,,
\ee
where $T_4=T/10^4$ K. The equilibrium temperature corresponds to both sides of
the equation vanishing, and has the value 
\be
T_{\rm eq} \approx 78\,{\rm K}\,
\left({n\over300\,{\rm cm}^{-3}}\right)^{1/3}\, ,
\label{tempequil} 
\ee
where we have taken $\epnuke=1\,{\rm keV}$.

In the discussion thus far, the nuclear reactions took place on sufficiently
long time scales that we did not need to consider the time evolution of $n_p$
and $n_n$ due to depletion of protons and neutrons.  If the reaction
$n(p,\gamma){\rm D}$ occurs when $n_p\neq n_n$, then the less abundant species
will exponentially vanish, while the more abundant species will asymptotically
approach a nonzero constant value.  For simplicity, we assume that the starting
densities of neutrons and protons are equal and evolve in the same manner.  We
denote the initial number density of either species as $n_0$. The
characteristic time scale for the populations of nuclei to change is the
inverse of the rate 
\be
\gamma = n_0 \sigmanp \approx 3\times10^{-17}\,{\rm s}^{-1}\,
\left({n_0\over300\,{\rm cm}^{-3}}\right) 
\sim {1 \over {\rm Gyr}} \,. 
\label{gammatime} 
\ee
The population of each nuclear species will thus decrease with time according
to the expression
\be
n(t) = {n_0 \over 1 + \gamma t} \,. 
\ee
The full differential equation for the time evolution of the (constant density)
cloud thus has the approximate form 
\be
{3 \over 2} {d \over dt} \left( kT \right) = 
- C T_4^3 + {\gamma \epnuke \over 1 + \gamma t}\,. 
\ee
Next we can write this evolution equation in dimensionless form by defining a
time scale $t_c=(3/2)kT_0/C$, where $T_0=10^4$ K (note that $t_c\sim0.67$ yr).
In terms of the dimensionless time $\tau=t/t_c$, the evolution equation becomes 
\be
{dT_4\over d\tau} = - T_4^3 + {B \over 1 + \Gamma \tau} \,,
\label{tempdiffeq} 
\ee
where $B=\gamma \epnuke/C \approx 4.8 \times 10^{-7}$ and where $\Gamma =
\gamma t_c \sim 6 \times 10^{-10}$.  Since the second term evolves on a much
longer time scale than the dimensionless time $\tau$ in the differential
equation, we can find an approximate solution by fixing the value of the second
term when solving the equation, and putting the time dependence back in
afterwards. We thus define 
\be
a = a(\tau)\equiv \left[{B\over 1+\Gamma\tau}\right]^{1/3} \,,
\ee
and integrate the differential equation to find the implicit solution
\begin{align}
\tau &= \int_{T_4}^1 {dT \over T^3 - a^3}\nonumber\\ 
&={1 \over 6a^2} 
\left\{ \log\left( {a^2 + a + 1 \over a^2 + a T_4 + T_4^2} \right)
+ 2 \log\left( {1-a \over T_4-a} \right) \right\}\nonumber\\
&+ {2 \sqrt{3} \over 6a^2} \left\{ 
\left[ \tan^{-1} \left( {a + 2 \over \sqrt{3}a} \right) - 
\tan^{-1} \left( {a + 2T_4 \over \sqrt{3}a} \right) \right] 
\right\}\,.
\end{align}
This form is rather cumbersome. We can also write the integrand as a series and
integrate term by term to obtain 
\be
\tau = \sum_{n=0}^\infty {a^{3n} \over 3n+2} 
\left[ T_4^{-(3n+2)} - 1 \right]\,. 
\ee
The first term gives us the simple form
\be
T_4(\tau) = {1 \over (1 + 2 \tau)^{1/2}} \,,
\ee
which describes the initial phase of evolution.  In contrast, the long term
evolution is given by assuming a quasi-steady-state solution for Eq.\
\eqref{tempdiffeq}, which implies 
\be
T_4(\tau) = {B^{1/3} \over (1 + \Gamma \tau)^{1/3}}\,.
\ee

These solutions indicate that the gas can cool relatively quickly (over time
scales measured in years) from an initial temperature of $T_0=10^4$ K down to
temperatures $T\sim T_{\rm eq}\sim80$ K.  Subsequent evolution and cooling
relies on the depletion of neutrons, which provide a nuclear heating source.
The relevant depletion time is $\sim1$ Gyr [see Eq.\ \eqref{gammatime}]. 

These results suggest that the formation of molecular clouds, and the
subsequent onset of star formation, will be only modestly affected by the
presence of nuclear reactions due to free neutrons. Heating due to these
reactions will slow the cooling of the gas material and thus delay star
formation. The depletion time is of order 1 Gyr, so that, after a delay of this
order, star formation could proceed unimpeded.  The nuclear reactions produce
deuterium, so that the stars that form will be enriched in deuterium relative
to those in our universe.

\subsection{Star Formation}\label{sec:starform} 

As the next stage of structure formation, molecular clouds produce small
centrally concentrated regions which constitute the actual formation sites for
individual stars \cite{sal87}. These structures, called molecular cloud cores,
slowly condense out of the larger cloud as they lose pressure support from both
turbulence and magnetic fields.  After the core regions reach a sufficiently
concentrated configuration, they undergo dynamic collapse, with a star forming
at the center of the collapse flow. A circumstellar disk forms around the star
and serves as a reservoir of angular momentum. This section shows that the
condensation phase of this sequence leads to the processing of most of the free
neutrons into deuterium. 

During the condensation phase, the density distribution of the molecular cloud
core can be described by a profile of the form \cite{adshu} 
\be
\rho(r,t) = {\Lambda \over 2\pi G t^2} {1 \over 1 + \xi^2} 
\qquad {\rm where} \qquad \xi = {r \over a|t|} \,. 
\label{coredense} 
\ee
In this expression, time $t$ is defined as the time {\it before} 
dynamic collapse begins, so that $t$ decreases as the core grows 
more concentrated. The dimensionless parameter $\Lambda$ specifies 
the initial overdensity of the region and is of order unity. The 
parameter $a$ is the effective transport speed in the gas. 

With the density distribution of Eq.\ \eqref{coredense}, molecular cloud cores
will remain optically thin until extremely short times before the onset of
collapse. These short times are not realized in practice, as the solution break
down once the time is shorter than $\sim10^4$ years. At this time before
collapse, the core experiences a rapid transition into its collapse state,
where this transient phase is no longer described by Eq.\ \eqref{coredense}.
The total optical depth, $\tau$, of the core is given by
\be
\tau = {\sigma_T \Lambda a \over 4\mpro G |t|} \approx 1.4\times10^{-3}
\left({|t| \over 1\,{\rm Myr}}\right)^{-1}\, ,
\label{taucore} 
\ee
where \mpro is the proton mass and we have taken $\Lambda a=0.3\,{\rm km/s}$.
The core thus remains optically thin until a time $t\sim1400$ yr before its
dynamic collapse, and hence for the entire evolutionary time of interest. 

The rate per unit volume at which neutrons are synthesized into deuterium has
the form 
\be
{dN_n \over dV dt} = n^2 \sigmanp\,. 
\ee
The total number of nucleons processed is larger by a factor of 2.  The total
conversion rate is determined by integrating over the volume of the cloud core
that will become the star. In this case, the reaction rate is a steeply
decreasing function of radius, so that we can ignore the outer boundary of the
core and integrate out to infinity, 
\be
{dN_n \over dt} = {\sigmanp \over \mpro^2} 
\int_0^\infty 4\pi r^2 dr \rho^2 = 
{\sigmanp \Lambda^2 a^3 \over 4 G^2 \mpro^2 |t|} \,. 
\label{nukerate} 
\ee
Including the factor of 2 to account for both the neutrons and the protons that
are processed, the total number of nucleons burned is given by the time
integral 
\be
\Delta N = 2\int_{t_0}^{t_f} {dN_n\over dt} dt = 
{\sigmanp \Lambda^2 a^3 \over 2 G^2 \mpro^2} 
\log\left[{t_0\over t_f}\right] \,.
\ee
The solution for the condensing core only holds up to a time $t_f\approx10^4$
yr before dynamic collapse. After this time, the core undergoes a transition
before rapidly approaching a well-defined collapsing state \cite{shu77}. The
initial time $t_0\approx1-10$ Myr is determined by the time when the solution
of equation (\ref{coredense}) first holds. The logarithmic factor is thus
$\sim\log10^3\sim7$, and number of converted nucleons is of order
\be
\Delta N \approx 8 \times 10^{56} \left({a \over 0.3\,{\rm km/s}}\right)^3
\approx 0.64 N_\odot\,, 
\ee
where $N_\odot$ is the number of nucleons in a solar mass star. Given that
roughly one third of the mass is already in deuterium (from BBN), this result
indicates that most of the remaining nucleons will experience nuclear reactions
during the contraction phase of the molecular clouds core that forms stars. Any
remaining free neutrons are then likely to be burned during the subsequent
dynamic collapse phase.  As a result, we expect stars to form with relatively
few free neutrons left, and to begin their evolution with a large deuterium
composition. 

The discussion thus far assumes that the energy produced by the nuclear
reactions has a negligible effect on the evolution of the condensing cloud
core. Equation \eqref{nukerate} specifies the number of reactions per unit time
integrated over the entire core structure.  The total luminosity (power)
generated by the core is thus given by
\be
L = \epnuke {d N_n \over dt} \,,
\ee
where $\epnuke$ is the energy per reaction that is retained by the gas.
Equation \eqref{taucore} shows that the core remains optically thin to the
radiation produced by the reactions, implying most of the energy (2.2 MeV per
reaction) is lost. Only the recoil energy is retained so that we expect
$\epnuke={\cal O}$(1 keV). The luminosity is thus given by 
\begin{align}
L &= {\epnuke \sigmanp \Lambda^2 a^3 \over 4 G^2 \mpro^2 |t|}\nonumber\\
&\approx 0.7 L_\odot \left({\epnuke\over1\,{\rm keV}}\right) 
\left({a\over0.3\,{\rm km/s}}\right)^3 
\left({t\over1\,{\rm Myr}}\right)^{-1}\,. 
\end{align}
For comparison, during the subsequent stage of dynamic collapse, 
the protostellar luminosities are $L\sim$ {\sl several} $L_\odot$, 
and are not large enough to affect the dynamics.\footnote[2]{Specifically, 
the luminosity is a fraction of the power scale $L_0=GM_\ast{\dot M}/R_\ast$,
where $M_\ast$ and $R_\ast$ are the mass and radius of the forming star at 
the given time and ${\dot M}\sim a^3/G$ is the rate at which mass falls into 
the central region.} 

The above considerations indicate that even though nuclear reactions
can process most of the free neutrons into deuterium during the phase
of core condensation, the energy generated has little effect on the
evolution. This finding may seem counterintuitive: Nuclear reactions
in our universe can power the Sun for $\sim10$ Gyr, whereas core
evolution takes place on the much shorter time scale of $\sim1$ Myr,
but both systems burn up comparable amounts of nuclear fuel.  Even
though the core processes essentially all of its nuclei at this faster
rate (by a factor of $\sim10^4$), each reaction provides only $\sim1$
keV of usable energy, compared to $\sim28$ MeV for each helium
nucleus produced in the Sun. The effective energy resources are thus
also different by a comparable factor ($\sim10^4$) so that the object
has about the same power ($L\sim1\,L_\odot$). This result holds only
because the core remains optically thin to the gamma rays produced by 
the reactions.

\section{Stellar Evolution}\label{sec:stars}

\subsection{Changes to the MESA package}

In this section we detail the changes we made to \mesa
\cite{MESA:2011,MESA:2013} to compute stellar evolution in the absence of the
weak interaction.  The primary reaction chain for \heiv synthesis in our
universe is the $pp$ chain, schematically given as
\begin{align}
  p+p&\rightarrow {\rm D}+\nue+e^+\label{eq:pp1}\\
  p+{\rm D}&\leftrightarrow \hethree+\gamma\label{eq:pp2}\\
  \hethree+\hethree&\leftrightarrow\heiv + 2p.\label{eq:pp3}
\end{align}
The reactions in Eqs.\ \eqref{eq:pp2} and \eqref{eq:pp3} are electromagnetic
and strong, respectively, and would be in operation in a weakless universe.
Eq.\ \eqref{eq:pp1} is a weak interaction and by definition is no longer
applicable.  As a result, we remove $p(p,\nue e^+){\rm D}$ from the nuclear
reaction network in \mesa, while preserving ${\rm D}(p,\gamma)\hethree$ and
$\hethree(\hethree,2p)\heiv$.

The BBN calculations in Table \ref{tab:yields} show that the primordial
composition can have significant contributions from ${\rm D}$.  Free protons
can capture on the ambient ${\rm D}$ to form \hethree, in line with Eq.\
\eqref{eq:pp2}.  Additionally, two ${\rm D}$ nuclei can interact with each
other to form larger nuclei through three channels
\begin{align}
  {\rm D}+{\rm D}&\leftrightarrow\hethree + n\label{eq:dd1}\\
  {\rm D}+{\rm D}&\leftrightarrow{\rm T} + p\label{eq:dd2}\\
  {\rm D}+{\rm D}&\leftrightarrow\heiv + \gamma.\label{eq:dd3}
\end{align}
All three of the reactions are in operation in a weakless universe.  Reactions
\eqref{eq:dd1} and \eqref{eq:dd2} are the principal means of ${\rm D}$
destruction whereas reaction \eqref{eq:dd3} is subsidiary.  To properly compute
the nucleosynthesis in stars, we must include all three reactions in our
nuclear reaction network, and in addition, we must include $n$ and tritium,
${\rm T}$, in our isotope list.

With the inclusion of free neutrons, we need to include other nuclear
reactions, for example, ${\rm T}({\rm D},n)\heiv$ and ${\rm T}({\rm T},2n)\heiv$.  Our
final nuclear reaction network includes all of the BBN reactions which involve
$A<5$ from Ref.\ \cite{SMK:1993bb}.  In addition, we include other reactions
which are not important in BBN but could be important with a high ${\rm D}$ mass
fraction, e.g., ${\rm D}({\rm D},\gamma)\heiv$ from Eq.\ \eqref{eq:dd3}.  Finally, our
nuclear reaction network includes reactions which synthesize \carbonxii and
\osixteen for completeness.  \carbonxii is the catalyst for the CNO cycle which
burns free protons into \heiv.  The CNO cycle relies on $\beta$ decays which
are not in operation in a weakless universe.  We do not include any part of the
CNO cycle in our calculations.

\subsection{Weakless stars}

We compute the deuterium-burning main sequence at zero metallicity for three
cases of the weakless universe as shown in Table \ref{tab:yields}:
$\eta =
10^{-9}, 10^{-10}, 10^{-11}$.
For all cases of $\eta$, we
fix $\np=1$.  The range of $\eta$ includes the value for our universe of
$6\times10^{-10}$ \cite{PlanckXIII:2015} and the value adopted by Ref.\
\cite{HKP:2006_weakless} of $4\times10^{-12}$, plus an intermediate value.

At still lower values of $\eta$, a negligible amount of primordial
nucleosynthesis occurs, and the universe consists almost entirely of free
protons and free neutrons, which convert to deuterium during star formation.
The evolution of these stars would be similar to the $\eta = 10^{-11}$ case,
but their lifetimes would be roughly twice as long. At higher values of $\eta$,
BBN produces a universe composed almost entirely of \heiv.

Notably, the only scenario that produces long-lived stars with Gyr lifetimes is
$\eta = 10^{-11}$. Therefore, we adopt this as the ``weakless universe'' where
not otherwise specified, including the discussion on habitability in Sec.\
\ref{sec:ChemEvo}. We compute three other stellar main sequences for comparison
to this weakless universe model. The weakless universe with metals is a model
for stars in a weakless universe that has undergone chemical evolution as
described in Section \ref{sec:ChemEvo}. This model adopts mass fractions of
$X_p = 0.01$ for free protons, $X_{\rm D} = 0.53$ for deuterium, $Y = 0.45$ for
\heiv, and $Z = 0.01$ for metals. ``Our universe'' is simply the main sequence
in our universe, also computed with metals. We compute this because it is most
recognizable on the Hertzsprung-Russell (H-R) diagram. Finally, we define a
``weak analog universe,'' which has a weak interaction, but it has the same
helium abundance as the weakless universe ($Y=0.4569$) and no metals to make
the closest possible comparison to the metal-free weakless universe.

We are unable to compute the very bottom end of the main sequence in a weakless
universe because the minimum stellar mass is 0.013 $M_\odot$, the
deuterium-burning limit, which the minimum mass computed by MESA is 0.03
$M_\odot$. There are also gaps in some of our computed main sequences because
MESA failed to converge.

\begin{figure}[htbp]
  \includegraphics[width=0.99\columnwidth]{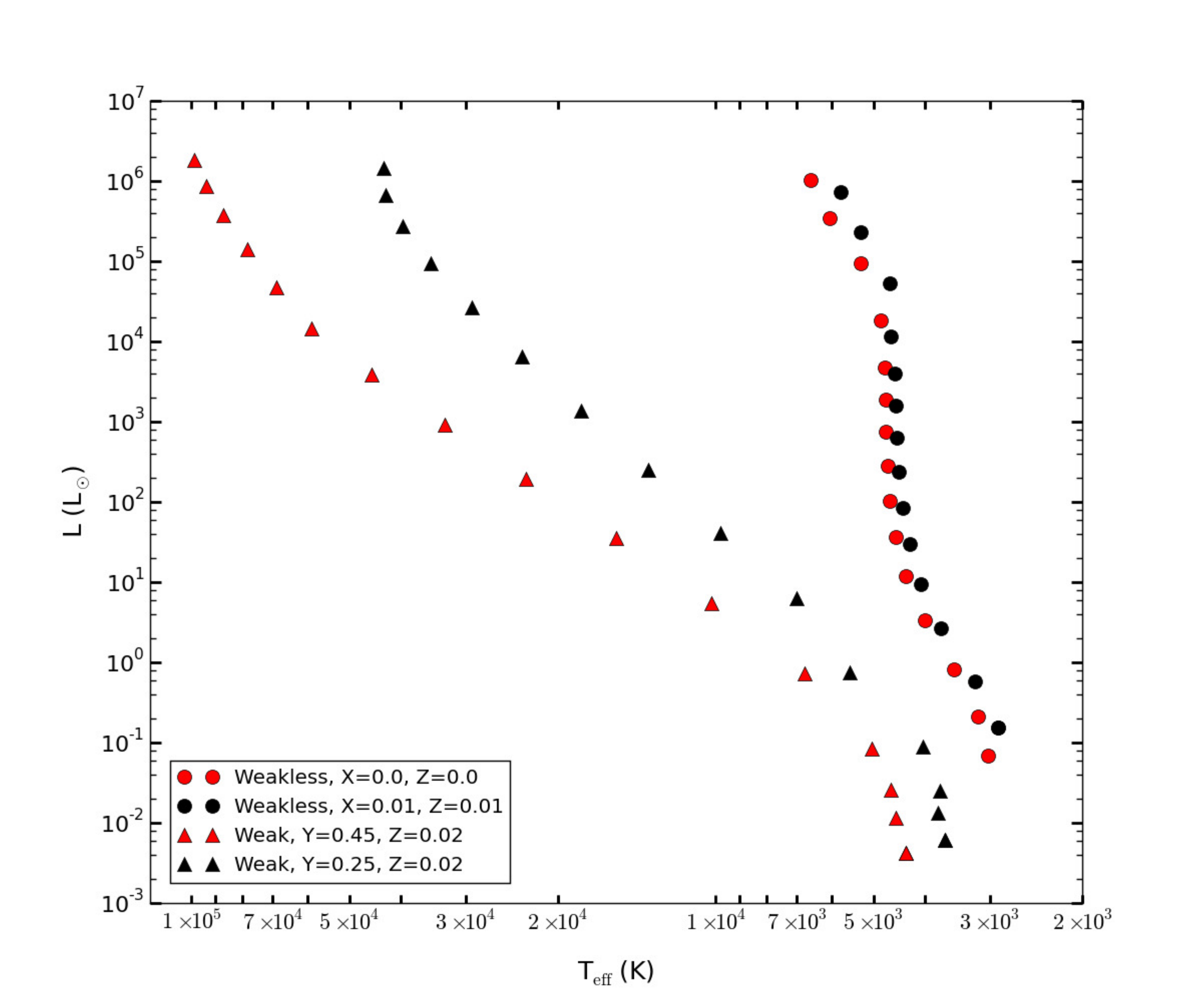}
  \caption{The stellar main sequence in our universe (black triangles), a
  ``weak analog'' universe (red triangles), and the weakless universe (circles)
  plotted on the H-R diagram. For weak universes, this is a $pp$-burning main
  sequence, while in the weakless universe, it is a deuterium-burning main
  sequence. For the weakless universe, we plot the main sequence with and without
  metals (black and red, respectively). The weak analog universe has no metals
  and the same helium fraction as the weakless universe.}
  \label{HRdiagram}
\end{figure}

We plot the stellar Zero-Age Main Sequence (ZAMS) from all four of these models
in Fig.\ \ref{HRdiagram} on the H-R diagram.  The start of hydrogen burning
defines the ZAMS.  Our universe is represented with black triangles, the weak
analog universe with red triangles, our adopted weakless universe with red
circles, and the weakless universe with metals with black circles. Note that
the main sequence in weak universes is a $pp$-burning main sequence, while in
weakless universes, it is a deuterium-burning main sequence.

Our universe produces the familiar main sequence on the H-R diagram, while a
more helium-rich universe produces hotter stars. In contrast, the main sequence
in a weakless universe falls mostly along the Hayashi track, lying vertically
over much of its length, staying near $\teff = 4.5\times10^3$ K for a wide range of
masses. Stars more massive than about 50 $M_\odot$ are bluer, and stars less
massive than 0.25 $M_\odot$ are redder. This occurs because deuterium burning
begins at a much lower central temperature and thus a much earlier time than
the $pp$-chain. The protostars descend down the Hayashi track as in our
universe, until deuterium burning reaches an equilibrium level, which occurs
while the star is still large and red. As a result, most stars in a weakless
universe appear as red giants. The lower metallicity and higher helium fraction
results in them being more orange, but they do become redder by 500-1000 K as
they age.

\begin{figure*}[htbp]
  \includegraphics[width=0.99\textwidth]{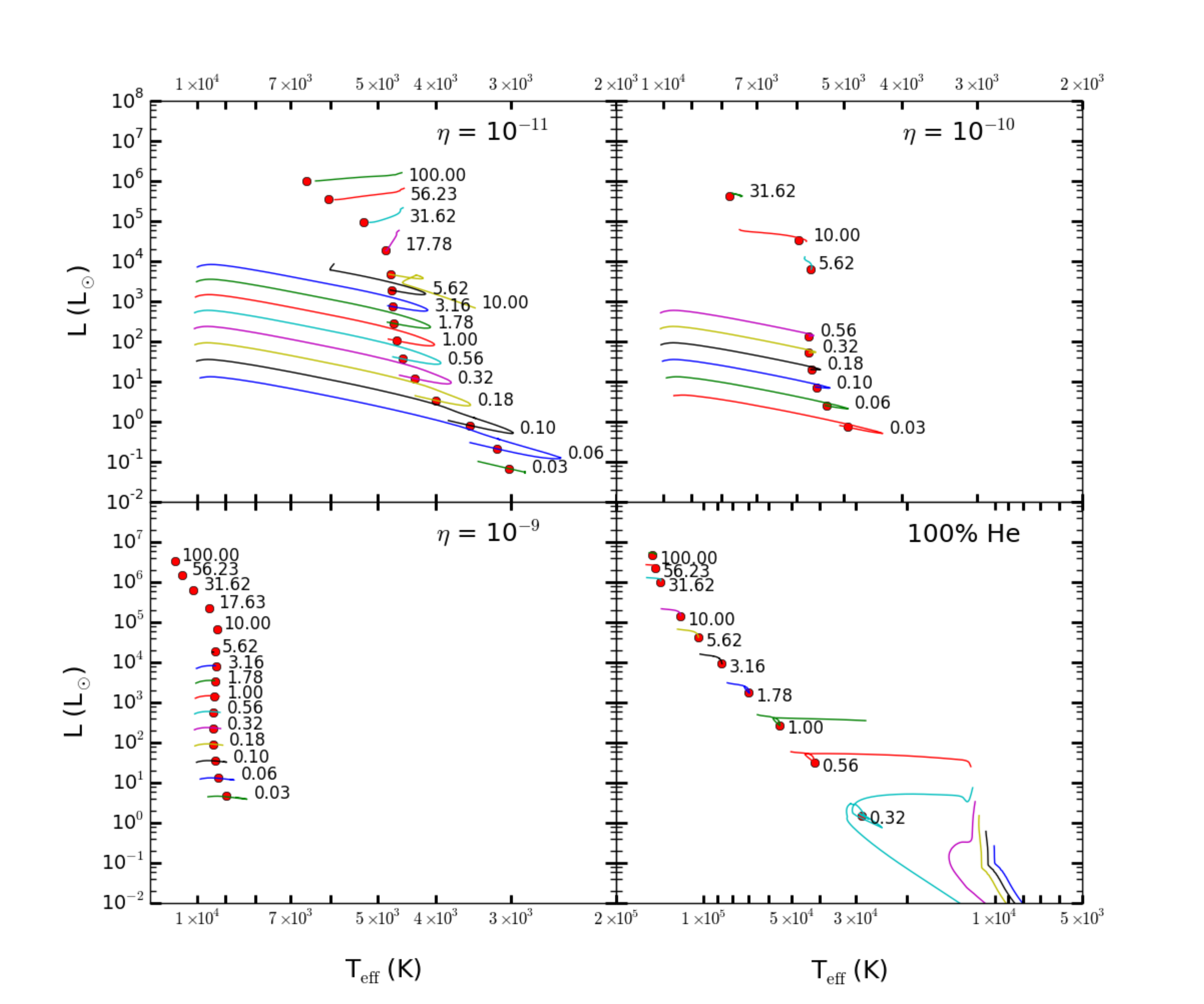}
  \caption{Evolutionary tracks of stars plotted on the H-R diagram in our three
  weakless universe cases and pure helium stars. Masses of individual tracks are
  labeled in solar masses. The tracks begin at an age of $10^5$ years 
  and end at the end of deuterium burning (helium burning for pure helium
  stars). The red dots denote the ZAMS for the weakless models, and
  the start of helium burning for the 100\% helium models.}
  \label{evotracks}
\end{figure*}

Figure \ref{evotracks} shows evolutionary tracks on the H-R diagram for our
stellar main sequences for all three of our weakless universe models through
the end of deuterium burning as well as a fourth model computed with 100\%
\heiv, which represents the limit as $\eta$ increases. We begin plotting the
tracks at an age of $10^5$ years, reflecting
the fact that it takes that long to make a protostar in our universe -- the
time to build up a protostar to its final mass and reach the birth line. This
is an aspect of history that is not captured by \mesa, which computes a
pre-main sequence model at the final mass and a very large radius, then lets it
contract. Nonetheless, we are confident in our results at the 10\% level
because models with different initial conditions tend to converge quickly and
because our models look very similar at 10$^4$ years and 10$^5$ years,
indicating they have reached such a quasi-equilibrium state.
A red dot on each track denotes the location in the H-R diagram when the star
reaches the ZAMS.  Some red dots do not appear to be on the tracks which
indicates that those stars burn deuterium during protostar formation.  We also
put a red dot on the tracks for the 100\% \heiv models which denote the start
of \heiv burning.

For all of our weakless models, but especially for the $\eta = 10^{-11}$ case,
the evolutionary tracks look significantly different from our universe. Instead
of moving upward and slightly blueward on the H-R diagram as they age, weakless
stars of less than about 5 $M_\odot$ make a large redward excursion and grow
significantly fainter early in their lives, then move back to near their ZAMS
position, followed by a much larger blueward excursion near the ends of their
lives. The redward excursion is 500-1000 K cooler and takes $\sim10\%$ of the
stars' main sequence lifetime to reach the redmost point. Returning to the ZAMS
position takes a further $\sim70\%$ of the main sequence lifetime, but the
stars have grown significantly brighter by that point. For the smallest stars
that are of most interest for habitability, their brightness increases by a
factor of $\sim 5$ over the main sequence lifetime, as opposed to $\sim 2$ for
Sun-like stars in our universe. The blueward excursion takes the remaining
$\sim20\%$ of the main sequence lifetime and stops at about $10^4$ K for a wide
range of masses as the deuterium fuel is exhausted, and the star enters a new
contraction phase ahead of helium burning.

Stars in cases with a higher value of $\eta$ have a much lower deuterium
fraction.  This means that they must have a higher core temperature and
contract further to reach an equilibrium state, and they will also have a
higher surface temperature due to the higher helium fraction.  Thus, the main
sequence is shifted blueward while the endpoint of deuterium burning remains
roughly the same: a nearly pure-helium star with an effective temperature of
$\sim10^4\,{\rm K}$.  The stars will follow similar, but much
shorter-lived evolutionary tracks.  For $\eta=10^{-10}$, the longest lived
stars live a few hundred Myr and typical effective temperatures are around
$6\times10^3\,{\rm K}$.  For $\eta=10^{-9}$, stars live only a few Myr at most
and have typical effective temperatures around $9\times10^3\,{\rm K}$, in both
cases moving blueward to the same endpoint.

The limiting case of pure helium stars will not occur in a weakless universe
even for very high baryon densities of $\eta=10^{-6}$ because the helium
fraction produced by BBN appears to approach a limit of $\sim98\%$.  However,
we still include them here for comparison purposes.  The evolutionary tracks
for these stars represent the helium burning phases.  These helium stars do not
``stall'' at a deuterium-burning main sequence and instead continue contracting
until they reach a helium-burning main sequence (the point at which the power
output from helium burning overwhelms that of hydrogen/deuterium burning) with
temperatures of $3\times10^4$ -- $15\times10^4\,{\rm K}$, while stars smaller
than $\sim0.3\, M_{\odot}$ fail to start helium burning at all.

\begin{figure}[htbp]
  \includegraphics[width=0.99\columnwidth]{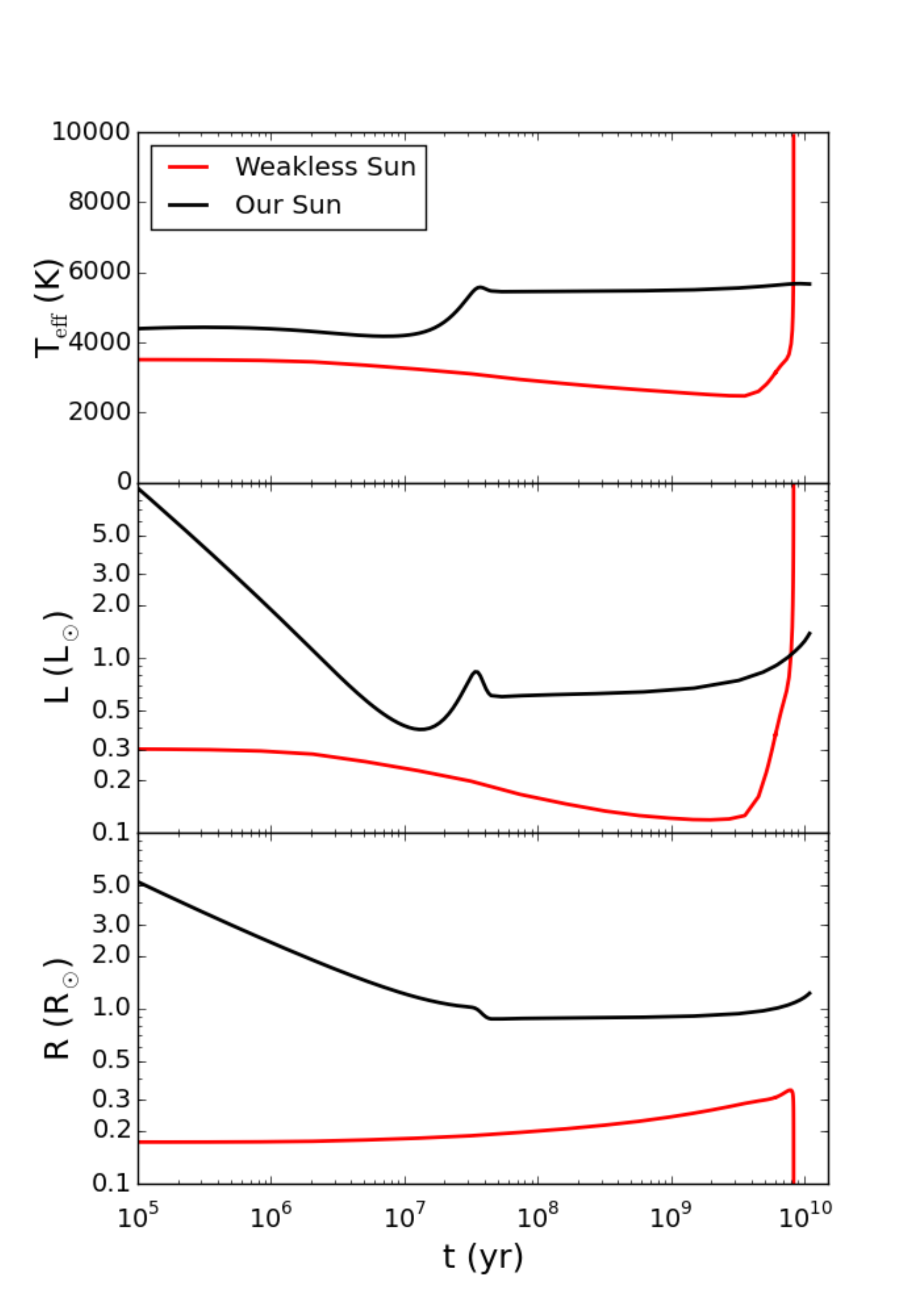}
  \caption{Comparison of the time evolution of temperature, luminosity, and
  radius of our sun and our adopted ``weakless sun'' with a mass of 0.056
  $M_\odot$ and $\eta = 10^{-11}$.}
  \label{timeevo}
\end{figure}

Of the cases we study, only small stars ($\lesssim 0.1 M_\odot$) in the $\eta =
10^{-11}$ case have the multi-Gyr lifetimes needed to have a large chance of
supporting habitable planets. For comparison purposes, we adopt a ``weakless
sun'' with a mass of 0.056 $M_\odot$ and a lifetime of 8.3 Gyr, and we plot its
temperature, luminosity, and radius evolution compared with our sun in Figure
\ref{timeevo}.

Unlike our sun, the weakless sun grows significantly fainter and redder over
the first Gyr of its life, decreasing in luminosity by a factor of 3. By
itself, this is not too different from M-dwarfs in our universe, but the
redward excursion that occurs on the main sequence (top-left panel of Fig.\
\ref{evotracks}) is still unusual. The weakless sun resembles a bright M-dwarf
with a temperature of $2.5\times10^3$ K, a luminosity of 0.1 $L_\odot$, and a
radius 0.15 $R_\odot$ during the period when its properties are most stable.

The other striking difference between the evolution of the weakless sun and our
sun is that despite its main sequence lifespan of 8.3 Gyr, the weakless sun
begins to take a sharp upturn in temperature and luminosity at an age of 3.6
Gyr, increasing in brightness by a factor of 10 by the end of deuterium
burning, whereas a solar-type star brightens much more slowly and more steadily
over its main sequence life. This shift corresponds to the blueward excursion
on the H-R diagram.

\begin{figure*}[htbp]
  \includegraphics[width=0.99\textwidth]{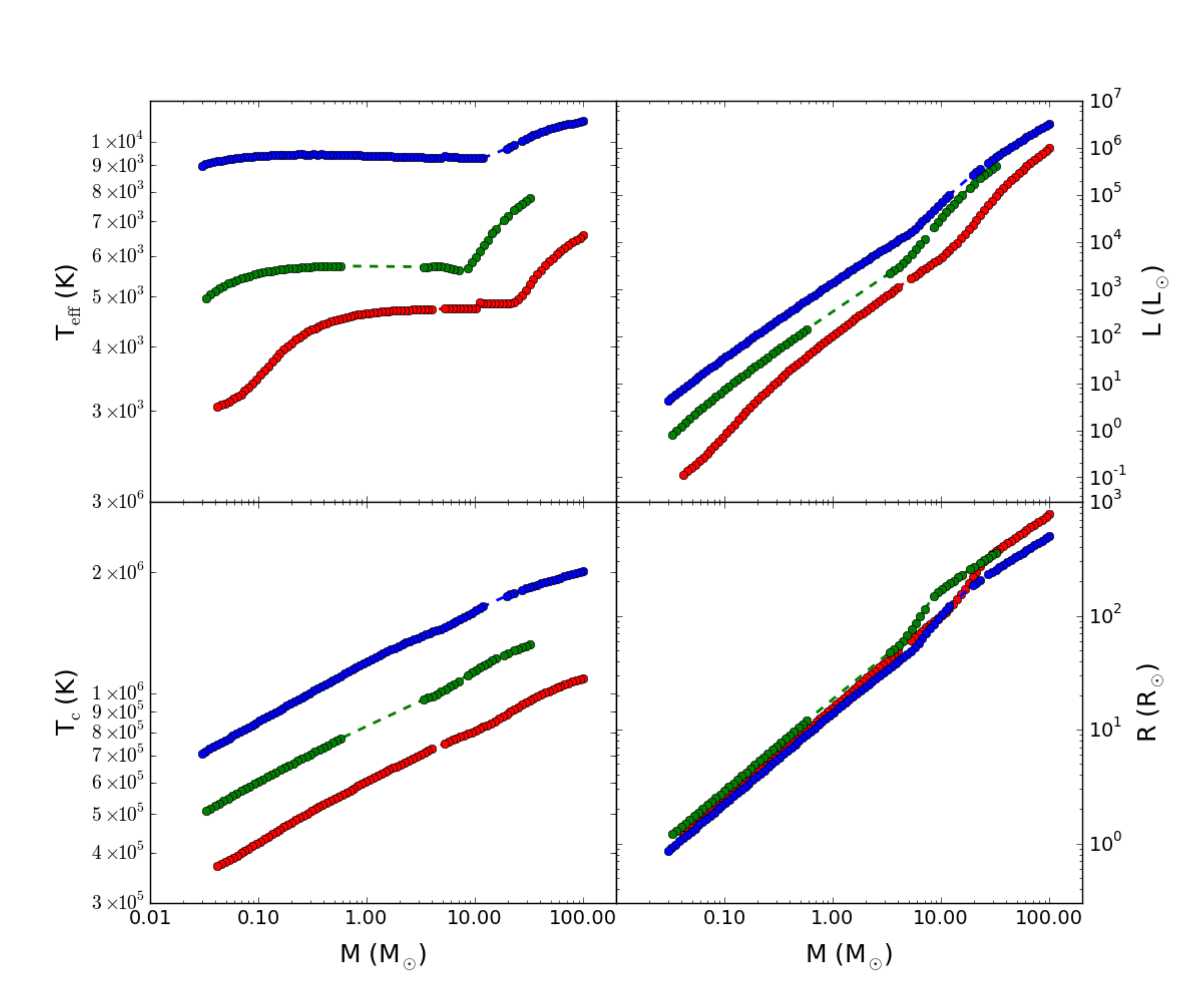}
  \caption{Plots of ZAMS stellar parameters versus mass for our three weakless
  universe models. Top-left: effective temperature. Top-right: luminosity.
  Bottom-left: core temperature. Bottom-right: radius.}
  \label{massplot}
\end{figure*}

Finally, we plot important ZAMS stellar parameters versus mass in Fig.\
\ref{massplot} for all three of our weakless universe cases, here computing a
much greater number of individual models. The top two panels, showing effective
temperature and luminosity, mirror the results we see in the H-R diagram. We
note that the mass luminosity relation for weakless stars have a much shallower
slope of $L\propto M^2$ compared with our universe, where the relation is
closer to $L\propto M^{3-4}$.

Because the temperature of weakless stars is nearly constant over a wide range
of masses, we expect to find a mass-radius relation of $R\propto M$, such that
radius scales linearly with mass. This is borne out by our plot of radius
versus mass in the bottom-right panel. This results in the smallest stars being
brighter, while the largest stars have a similar luminosity to those in our
universe on the order of $10^6 L_\odot$.

The bottom-left panel of Fig.\ \ref{massplot} plots the central
temperature versus mass for our three weakless universe cases and shows
significant variation with mass. The deuterium-burning temperature is usually
described at $10^6$ K.  However, the lowest-mass stars in our $\eta = 10^{-11}$
case have central temperatures approaching $3\times10^5$ K, much lower than
expected. It may be that the much higher deuterium concentration and the higher
central density make up for the lower temperature in the reaction rate, or that
following the stellar evolution over Gyr time scales implies that lower
reaction rates can be significant where they would not be in our universe.
In any case, the central temperature also has a very consistent
power law relation with stellar mass of $T_c\propto M^{0.14}$.


The mass-luminosity relationship shown in the upper right panel of Fig.\
\ref{massplot} has the power-law form $L\propto M^2$.  Although a detailed
derivation of this relation is beyond the scope of this paper, we can
understand this finding in approximate terms. Firstly, we note that this
scaling law is intermediate between the mass-luminosity relation found for low
mass stars in our universe ($L \propto M ^4$) and that for high mass stars ($L
\propto M$). The weakless stars under consideration here have properties in
common with both low-mass and high-mass stars.  Weakless stars of low mass are
brighter than those in our universe, but objects at the high mass end of the
distribution are somewhat dimmer. This compression of the luminosity range
leads to the intermediate slope of the mass-luminosity relationship.  This
finding can also be understood through the following approximate derivation. 

Using order of magnitude scaling laws \cite{hansen,kippenhahn}, we can write
the central pressure of the star in terms of the stellar mass and radius,
\be\label{centralp}
P_c \sim {GM^2 \over R^4} \,,
\ee
where dimensionless constants of order unity are suppressed.  Using the ideal
gas law to evaluate the pressure, we find 
\be\label{tvspotential}
kT \sim {GM\mu \over R} \,, 
\ee
where $\mu$ is the mean molecular weight and all quantities are evaluated at
the stellar surface.  In writing Eq.\ \eqref{tvspotential}, we ignore
dimensionless constants so that we can explore the scaling relationships.  The
dimensionless constants could be orders of magnitude different between the
center in Eq.\ \eqref{centralp} and the surface in \eqref{tvspotential}.  Next,
we note that the photospheric temperature is nearly  constant across the entire
stellar mass range, as indicated by the nearly vertical main sequence in Fig.\
\ref{HRdiagram} and the effective temperature plotted in the upper left panel
of Fig.\ \ref{massplot}.  This trend of a low, nearly constant surface
temperature is much like the behavior of stars ascending the red giant branch
in our universe (as noted earlier, these weakless stars have much in common
with red giants). In the case of red giants, as the stellar envelope expands
and the surface temperature decreases, the opacity of the photosphere
eventually increases due to contributions from H$^{-}$ ions. In addition, when
the photosphere reaches a minimum temperature of $T\sim5\times10^3$ K, the
outer layers of the star become fully convective.  Luminosity is efficiently
carried out of the star and prevents further lowering of the surface
temperature. As a result, red giants move almost vertically up the H-R diagram
(at nearly constant temperature).  The behavior in the top right panel of Fig.\
\ref{massplot} is thus the weakless analog of the well-known Hayashi forbidden
zone, which arises in pre-main-sequence evolution \cite{hayashione} and in red
giants \cite{hayashitwo}.  Weakless stars behave in a similar manner to these
two stellar states from our universe. As a result, Eq.\ \eqref{tvspotential}
implies that $M \propto R$.  The stellar luminosity is given by 
\be
L = 4\pi R^2 \sigma T^4 \sim R^2 \sim M^2\,,
\ee 
where $\sigma$ is the Steffan-Boltzmann constant and the final approximate
equalities assume that the surface temperature is constant.

\section{Chemical Evolution and Habitability}
\label{sec:ChemEvo}

Of the cases we consider, only the $\eta = 10^{-11}$ case produces stars with
Gyr lifetimes. In the $\eta = 10^{-10}$ case, we can extrapolate from our
results that a star at the deuterium-burning limit would have a lifetime of
$\sim0.6$ Gyr. However, in the $\eta = 10^{-11}$ case, stars of 0.013-0.10
$M_\odot$ would be similar to M-dwarfs and would have long lifetimes of 3-30
Gyr, long enough for complex life to develop on orbiting planets with liquid
water.

There are a few differences in weakless stars that impact habitability. Most
significantly, the weakless sun only remains in conditions stable enough to
support life for about 30\% of its main sequence lifetime, compared with 60\%
or more for a solar-type star, making them significantly less hospitable to
life. Additionally, the habitable zone of our weakless sun model would be
at roughly 0.3-0.5 AU, which is outside the tidal locking radius
\cite{2014ApJ...787L...2Y}, so tidal locking is not a concern.

The habitability of the weakless universe also depends on the presence of the
elements needed to make organic compounds. Chemistry in a weakless universe
would be nearly identical to our own, so we postulate that life would require,
at a minimum, carbon and oxygen. Additionally, life requires materials that can
form planets, although the possibility of water worlds means that this does not
necessarily require new elements.

The chemical evolution of a weakless universe is much more speculative once
later-generation stars form with a nonzero metallicity because the nuclear
reactions involved are not well-studied.  However, the initial generation of
stars will disperse their heavy elements via two primary mechanisms: red giant
winds and Type Ia supernovae.  Core-collapse supernovae (notably, the primary
source of oxygen in our universe) fail to explode because of the lack of
neutrinos, with the entire star collapsing to a degenerate remnant.  If the
star has undergone dramatic mass loss, it could collapse to a ``nucleon star''
analogous to a neutron star in our universe.  The maximum mass of a neutron
star is computed as 2.01 -- 2.16 $M_\odot$ \cite{2017arXiv171100314R}.
However, a nucleon star is composed of both protons and neutrons, giving the
degenerate matter twice as many degrees of freedom, so the maximum mass of a
nucleon star could potentially be as high as $4.32\,M_\odot$.  Nonetheless,
most massive stars will collapse directly to a black hole since they will not
undergo enough mass loss to form such a nucleon star.

Asymptotic Giant Branch (AGB) winds from high-mass stars dredge up triple-alpha
products from the cores of stars and disperse the largest amount of carbon in
our universe \cite{1999A&A...342..426G,2000ApJ...541..660H}.  The metals in
these winds consist mostly of carbon, but they also include some oxygen.
Reference \cite{2000A&A...361L...1P} estimates that the triple-alpha process
produces \oxvi at a rate of $\sim7\%$.

Meanwhile, Ref.\ \cite{1999ApJS..125..439I} estimates yields of Type Ia
supernovae in eight different models, resulting in $^{56}$Ni yields between
$35\%$ and $90\%$, most likely near the upper end of that range. Most of the
remaining ejecta is composed of $^{54}$Fe and $^{58}$Ni. Other alpha-process
elements, $^{28}$Si, $^{32}$S, $^{36}$Ar, and $^{40}$Ca are also produced in
significant amounts at the same order of magnitude as their solar abundances.
Without beta decay and with a \np ratio near one, the most important products
of Type Ia supernovae in a weakless universe are most likely $^{56}$Ni and
$^{52}$Fe, both of which will be stable. Thus, the second generation of stars
will form with ``nickel peak elements'' and carbon and oxygen, the necessary
elements to form terrestrial planets and life. These planets will have a very
iron-rich, Mercury-like composition, but will also have carbon and oxygen.

One other problem for habitability is the very high carbon-to-oxygen ratio,
which would seem to suppress the formation of water. However, Ref.\
\cite{2012ApJ...757..192J} suggests that in the reducing environment present at
high C/O ratios, as much as $10\%$ of the oxygen could still bind into water
instead of carbon monoxide, so this is also not necessarily a barrier to life
forming.

Many metals produced in stars will undergo further processing in the ISM. Just
as free protons combine with free neutrons to form deuterium during star
formation, these metals will also combine with free neutrons to form
neutron-rich isotopes. Any neutron capture reactions with a cross section
similar to or greater than that of free protons ($\sigma v = 7.3\times 10^{-20}
\,{\rm cm}^3\,{\rm s}^{-1}$) will occur in significant amounts, essentially
resulting in an $s$-process during star formation. Without beta decay, this
neutron process could theoretically continue all the way to the neutron drip
line, but neutron capture cross sections become small compared with free
protons for neutron-rich isotopes, so this is unlikely in practice.

The neutron capture cross sections have not been measured for all of the
isotopes in question, but it is known that they are several times greater than
those for free protons from $^{58}$Ni through $^{64}$Ni and for $^{54}$Fe through
$^{58}$Fe. Thus, each atom of nickel peak elements in a star-forming core will
absorb on the order of ten neutrons during star formation. Furthermore, the
same is true of calcium; argon atoms will absorb at least five neutrons each,
and sulfur atoms two or three. $^{12}$C, $^{16}$O, and $^{28}$Si will undergo
very little such processing due to small neutron capture cross sections.
Because the \np ratio is identical to one, this will result in
second-generation stars forming with a slight overabundance of free protons.
Based on solar abundances \cite{2010ASSP...16..379L}, the abundance of free protons
will probably be on the order of $1\%$.

Weakless stars will not have a CNO cycle, which depends on beta decay. Instead,
any free protons will contribute to what Ref.\ \cite{HKP:2006_weakless} call
``proton-clumping'' reactions, in which protons are added to carbon and oxygen
nuclei, which are stable up to the proton drip line, essentially resulting in
an ``$sp$-process'' in the star.
For \carbonxii, the proton drip line limits these reactions to
\begin{align}
  p + {}^{12}{\rm C} &\rightarrow {}^{13}{\rm N} + \gamma, \\
  p + {}^{13}{\rm N} &\rightarrow {}^{14}{\rm O} + \gamma,
\end{align}
and similarly for $^{16}{\rm O}$
\begin{align}
  p + {}^{16}{\rm O} &\rightarrow {}^{17}{\rm F} + \gamma, \\
  p + {}^{17}{\rm F} &\rightarrow {}^{18}{\rm Ne} + \gamma.
\end{align}
However, if conditions are hot enough to overcome the Coulomb repulsion between
protons and larger nuclei, other reactions are possible
\begin{align}
  4p + {}^{28}{\rm Si} &\rightarrow {}^{32}{\rm Ar}, \\
  4p + {}^{32}{\rm S} &\rightarrow {}^{36}{\rm Ca}, \\
  2p + {}^{36}{\rm Ar} &\rightarrow {}^{38}{\rm Ca}, \\
  4p + {}^{40}{\rm Ca} &\rightarrow {}^{44}{\rm Cr}.
\end{align}
Interestingly, there is no reaction chain that has nitrogen or fluorine as an
endpoint (every stable isotope of nitrogen and fluorine can become a stable
isotope of oxygen and neon, respectively, by adding a proton), or indeed most
odd-atomic-number elements. A weakless universe might therefore have a nitrogen
abundance an order of magnitude lower than our universe, like other odd-atomic
number elements, and a greater neon abundance, similar to carbon and oxygen in
our universe. The lower nitrogen abundance could have implications for life,
but on the other hand, these reactions do increase the abundance of oxygen,
potentially solving the problem of the high C/O ratio.  Further analysis of
proton-clumping reactions would be needed to determine how these abundances
evolve over time.

If carbon-based life arises in a weakless universe, it will have one further
difference from our universe in that nearly all of the hydrogen is in the form
of deuterium, and thus, nearly all of the water will be heavy water, D$_2$O. On
Earth, most plants and animals will die if roughly 50\% of the water in their
bodies is replaced with D$_2$O \cite{NYAS:NYAS736}. The first place to look for
the reason for this effect would be cellular respiration. All eukaryotic life
on Earth, including humans, derives energy by pumping protons across a membrane
via proton pump proteins to create an electrochemical potential which then
powers adenosine triphosphate (ATP) synthase proteins to form energy-storing
ATP molecules. However, this process is not significantly affected by the
introduction of heavy water; the proton pumps appear to work just as well as
deuteron pumps \cite{NYAS:NYAS736}.

The exact cause of the toxicity of heavy water remains uncertain, but it is
believed to be due to the altered strength of hydrogen bonds (intermolecular
forces between polar molecules) involving deuterium atoms \cite{KBD:1999_D2O}.
The hydrogen bonds between adjacent water molecules have a bond energy of 21 kJ
mol$^{-1}$, and deuteration increases the strength of these bonds on the order
of $10\%$ \cite{F19757100980}.  Because protein folding is determined in large
part by hydrogen bonds, any change in their strength can dramatically impair
cellular processes.  It is believed that the most important toxic effect
resulting from this is damage to fast-dividing cells, similar to the symptoms
of cytotoxic poisoning and radiation poisoning \cite{KBD:1999_D2O}.

In a weakless universe, where the majority of hydrogen is deuterium, enzymes
and biochemical reactions would have to adapt to the strength of
deuterium-based hydrogen bonds and the other quantum chemical properties of
deuterium. But this is no greater an evolutionary challenge than producing
biochemistry based on light hydrogen, so we do not consider it an impairment to
life.

One other factor deserves note: the abundance of many elements that are crucial
to biochemistry on Earth in ionic form is much lower in a weakless universe
because they are primarily produced by core-collapse supernovae. Sodium in
particular would be nearly nonexistent, and chlorine would also be depleted by two
orders of magnitude relative to our universe. Based on the yields of Type Ia
supernovae, the most common salt that is soluble in water is likely to be
magnesium sulfate (Epsom salt in its hydrate form), which is the second largest
component of sea salt on Earth.

\section{Conclusion}\label{sec:conclude} 

\subsection{Summary of Results}\label{sec:summary} 

The overarching result of this work is that universes in which the
weak interaction is absent can remain viable. This builds upon the
original proposal of a weakless universe \cite{HKP:2006_weakless} and is
largely consistent with that scenario. More specifically, our main
results can be summarized as follows.

We have studied the epoch of Big Bang Nucleosynthesis in detail,
exploring a wide range of baryon-to-photon ratio $\eta$ and over the
full range of possible initial neutron-to-proton ratios (Section
\ref{sec:bbn}). In order for the universe to avoid the overproduction
of helium, which would lead to a shortage of hydrogen, the value of
$\eta$ must be smaller than that of our universe by a factor of
$\sim100$. For the working range of parameters space, universes emerge
from the BBN epoch with roughly comparable abundances of protons, free
neutrons, deuterium, and helium. 

The main difference between a weakless universe and ours is the
presence of a substantial admixture of both free neutrons and
deuterium. However, the formation of galaxies (Section
\ref{sec:galaxy}) is largely unchanged. On these large spatial (and
mass) scales, the densities are too low for the nuclear composition to
play a role. The formation of substructure within galactic disks, such
as molecular cloud complexes where stars form (Section
\ref{sec:clouds}), is only modestly affected.  Some nuclear reactions
of the free neutrons can occur at cloud densities, and the resulting
heating can delay the onset of star formation, but the energy
injection is not sufficient to destroy the clouds. As substructures
within the clouds condense further (Section \ref{sec:starform}),
nuclear reactions involving the free neutrons and protons process
essentially all of the free neutrons into deuterium. As a result, the
free neutrons are used up before they are incorporated into stars.


Finally, we have considered stars and stellar evolution in universes without
the weak interaction (Sec.\ \ref{sec:stars}). For the low value of $\eta$ we
primarily consider here, stars begin their evolution with much greater amounts
of deuterium than in our universe. In the absence of the weak interaction, the
standard $pp$-chain and CNO cycle for hydrogen fusion are no longer operative,
and stars are powered by the strong interaction via deuterium burning (roughly
analogous to the scenario where diprotons are stable and stars must burn
exclusively through the strong interaction \cite{2015JCAP...12..050B}). These
stars resemble red giants and have proportionately greater luminosities and
shorter lifetimes, but the smaller minimum mass of deuterium-burning stars
means that stars with Gyr lifetimes are still possible. The lightest possible
stars more closely resemble late-K or early-M dwarfs and can live for up to
$\sim 30$ Gyr. These stars also show more variation in luminosity over their
main sequence lifetimes than stars in our universe.

After deuterium is processed into helium, the subsequent nuclear reactions that
take place via the strong interaction proceed in much the same way as in our
universe.  The weakless universe is similar to our universe with large
abundances of alpha-elements.  However, the dispersal mechanisms in a weakless
universe are limited to AGB winds for \carbonxii and Type Ia supernovae for the
other alpha-elements.  Projecting the chemical evolution of a weakless universe
over multiple generations of stars suggests some differences from our universe,
such as an overabundance of carbon and neon and an underabundance of nitrogen
and elements heavier than nickel, but these differences are not enough to
preclude either planet formation or the development of organic chemistry.

\subsection{Discussion}\label{sec:discuss} 

In addition to indicating that hypothetical universes without weak interactions
remain viable, the results of this work provide insight into the workings of
our own universe. By considering such scenarios, we can understand how far
trends in our universe can be taken.  We understand BBN in our universe such
that we can predict the primordial abundances to high precision (with the
exception of \livii \cite{2016JPhCS.665a2001C}).  Much current research done on
SBBN is in computing higher-order effects detectable in future experiments
\cite{cmbs4_science_book,EELT,GMT,2015RAA....15.1945S}.  However, the
relationship between $Y_P$ and \np is approximated to high accuracy as
$Y_P\simeq2(\np)/(1+\np)$.  This expression results from the assumption that
all of the neutrons (to one part in $10^5$) are incorporated into \heiv.  It
has traditionally been used in the context of an evolving \np ratio [given by
the reactions in Eqs.\ \eqref{eq:np1} -- \eqref{eq:np3}] and a baryon number
within an order of magnitude of that given by the CMB temperature power
spectrum.  In this work, we expanded the parameter space to encompass both a
larger portion of the range $0<\np<1$ and the new range $\np>1$.  We have found
that the expression $Y_P\simeq2(\np)/(1+\np)$ is valid over the larger range of
$0<\np<1$, and furthermore can be generalized with Eq.\ \eqref{eq:yp_np} to
extend to $\np>1$.  Although it is physically reasonable that with fewer than
$50\%$ protons, there will be less \heiv, what is surprising is the degree of
symmetry in Fig.\ \ref{fig:yp}.  This indicates that our weakless nuclear
reaction network --- identical to that of SBBN sans the weak interaction rates
--- isolates the baryons into \heiv and free nucleons independent of isospin.
An interesting scenario would be to extend the BBN network to include heavier
nuclei and more neutron capture reactions in the case of $\np>1$.  Such an
environment (high entropy and $\np>1$) could be conducive to $r$-process
nucleosynthesis, where there is no analog for $\np<1$ and hence an asymmetry.
The $r$-process relies on $\beta$-decays of neutron-rich nuclei, so our
argument would have to be applied to a class of universes where the weak
interaction is indeed present.  This thought experiment would test whether the
symmetry in Fig.\ \ref{fig:yp} is the result of a limited network, or something
more fundamental in the nuclear properties and interactions of the light
nuclides.  We note that this discussion is predicated on $\eta$ being large
enough so that out-of-equilibrium nuclear reactions can occur at low enough
temperatures.

While a weakless universe produces all of the elements necessary for life, the
relative habitability of such a universe compared with our own depends on
several factors such as the C/O ratio that are a result of chemical evolution.
In this paper, we have examined the most important effect on stellar evolution
by replacing the $pp$-chain and CNO cycle with a strong deuterium-burning
reaction. The next step would be to determine the reaction rates of
proton-clumping reactions and incorporate them into the nuclear network of
\mesa, which would make it possible to determine the stellar yields of carbon,
nitrogen, oxygen, neon, and possibly several heavier elements. These yields
would provide clearer insights into the range of possible chemical makeup of
life in a weakless universe.

Finally, the results of this paper suggest that the inclusion of nuclear
processing due to the presence of free neutrons alters the evolution of
galaxies, star formation, and stellar evolution to a moderate degree. The
changes are neither negligible nor dominant. In particular, the presence of
free neutrons does not prevent a universe from becoming habitable. In some
sense, this intermediate result can be understood on energetic grounds. The
presence of stable neutrons allows for fusion to take place in the interstellar
medium (rather than having all nuclear reactions take place in stellar
interiors) and produce deuterons, which have a binding energy of 2.2 MeV. For
comparison, the production of helium-4, with a binding energy of 28 MeV,
provides much of the energy for galaxies, so that the new energy source
represents an 8\% effect.

\acknowledgments 

We thank Juliette Becker, George Fuller, Lillian Huang, and Michael Meyer for
useful conversations.  This work was supported by the John Templeton foundation
through Grant ID55112: {\it Astrophysical Structures in other Universes}, and
by the University of Michigan.  Computational resources and services were
provided by Advanced Research Computing at the University of Michigan.  We
thank the referees for their comments.

\bibliography{master}

\end{document}